\documentclass{JHEP3}

\usepackage{epsfig}
\usepackage{amsbsy}
\usepackage{varioref}

\newcommand{\beq}{\begin{equation}}
\newcommand{\eeq}{\end{equation}}
\newcommand{\bea}{\begin{eqnarray}}
\newcommand{\eea}{\end{eqnarray}}
\newcommand{\nn}{\nonumber}

\def\eqn#1{Eq.~(\ref{#1})}
\def\eqns#1#2{Eqs.~(\ref{#1}) and~(\ref{#2})}
\def\eqnss#1#2{Eqs.~(\ref{#1})-(\ref{#2})}
\def\fig#1{Fig.~{\ref{#1}}}
\def\vfig#1{Fig.~{\vref{#1}}}
\def\sec#1{Section~{\ref{#1}}}
\def\app#1{Appendix~\ref{#1}}
\def\tab#1{Table~\ref{#1}}

\def\spa#1.#2{\left\langle#1\,#2\right\rangle}
\def\spb#1.#2{\left[#1\,#2\right]}
\def    \sapp#1#2#3#4{{\langle #1 | (#2+#3) |#4  \rangle} }

\def\bentarrow{\:\raisebox{1.3ex}{\rlap{$\vert$}}\!\longrightarrow}

\def\id{{\rm id}}
\def\i{{\rm i}}
\def\d{{\rm d}}

\def\vep{\varepsilon}
\def\sgg{s_{\gamma^*\gamma^*} }
\def\ord{{\cal O} }
\def\cA{{\cal A}}
\def\cP{{\cal P}}
\def\lra{\leftrightarrow}
\def\ellb{{\bar \ell}}
\def\eb{{\bar e}}
\def\ib{{\bar i}}
\def\qb{{\bar q}}
\def\Qb{{\bar Q}}
\def\epem{e^+e^-}

\newcommand\sss{\scriptscriptstyle}
\newcommand\as{\alpha_{\sss S}}
\newcommand\aem{\alpha_{\rm em}}
\newcommand\gs{g_{\sss S}}

\newcommand\fverb{\setbox\pippobox=\hbox\bgroup\verb}
\newcommand\fverbdo{\egroup\medskip\noindent%
                        \fbox{\unhbox\pippobox}\ }
\newcommand\fverbit{\egroup\item[\fbox{\unhbox\pippobox}]}
\newbox\pippobox

\title{The contribution of the four-parton final states
to $\boldsymbol{\gamma^*\gamma^* \to}$ hadrons}

\author{Vittorio Del Duca\\
Istituto Nazionale di Fisica Nucleare, Sez. di Torino\\
via P. Giuria, 1 - 10125 Torino, Italy\\
        E-mail: \email{delduca@to.infn.it}}

\author{Fabio Maltoni\\
Dept. of Physics, University of Illinois at Urbana-Champaign\\
Urbana, IL\ \ 61801, USA\\
        E-mail: \email{maltoni@uiuc.edu}}

\author{Zolt\'an Tr\'ocs\'anyi\thanks{Sz\'echenyi fellow of the Hungarian
Ministry of Education}\\ University of Debrecen and\\
Institute of Nuclear Research of the Hungarian Academy of Sciences\\
H-4001 Debrecen, PO Box 51, Hungary\\
        E-mail: \email{zoltan@zorro.atomki.hu}}

\abstract{In the analysis of the total cross section for the
$\gamma^*\gamma^*\to$ hadrons process, we include the four parton
final states, which are part of the $\ord(\as^2)$ corrections.
The four-parton final states contain the diagrams with gluon exchange in the
crossed channel, which constitute the leading order of the BFKL resummation.
We show that the diagrams with gluon exchange in the
crossed channel play an important role in the large $Y$ region, however
their contribution to the cross section must be evaluated exactly.
In fact, the high-energy limit, which constitutes the kinematic framework
of the BFKL resummation, is not sufficiently accurate at
LEP2 energies. The inclusion of the diagrams with gluon exchange in the
crossed channel reduces the discrepancy between the theory and the LEP2 data
collected by the L3 Collaboration, but the data still lie above the theory,
even allowing for a large scale uncertainty in the theory.
Thus, in order to describe accurately the data
for $\gamma^*\gamma^*\to$ hadrons in the large $Y$ region,
corrections of an order higher than $\ord(\as^2)$ seem to be necessary.
}

\preprint{{~DFTT 7/2002}\\{~hep-ph/0202237}}

\begin{document}

\section{Introduction}
\label{sec:intr}

Strong interaction processes, characterised by a large kinematic scale,
are described in perturbative QCD by a fixed-order expansion of the
parton cross section in $\as$, complemented, if the scattering process is
initiated by strong interacting partons, with the Altarelli-Parisi evolution
of the parton densities.
However, in kinematic regions characterised by two very different hard
scales, a fixed-order expansion might not suffice: large logarithms of the
ratio of the kinematic scales appear, which may have to be resummed.
In processes where the centre-of-mass energy $s$ is much larger than the
typical momentum transfer $t$, the sub-process which features gluon
exchange in the crossed channel, and that usually appears at $\ord(\as^2)$,
tends to dominate over the other sub-processes. That sub-process constitutes
the leading-order term of the BFKL equation, which is an equation for the
Green's function of gluon exchanged in the crossed channel.
The BFKL equation~\cite{Balitsky:1978ic,Kuraev:1976ge,Kuraev:1977fs}
resums the logarithms of type $\ln(s/|t|)$.

Over the last decade,
several observables, like the scaling violations of the
$F_2$ structure function~\cite{Aid:1995rk,Forte:1996xv},
forward-jet production in DIS~\cite{Aid:1995we}$-$\cite{Orr:1998tf},
dijet production at large rapidity
intervals~\cite{Abachi:1996et}$-$\cite{Andersen:2001kt},
and $\gamma^* \gamma^* \to$ hadrons in $\epem$
collisions~\cite{Acciarri:1999ix}$-$\cite{Bartels:1996ke}
have been proposed in the literature as candidates for the
detection of the BFKL evolution, and have been measured and analysed
as functions of observables, which aim to single out
large logarithms of type $\ln(s/|t|)$.
However, from a phenomenological point of view, in order to claim
detection of BFKL gluon radiation in a given process in an unambiguous way,
we must rule out any explanation of that process in terms of a fixed order
expansion, or in terms of a different resummation.
Thus, in order to make a sound BFKL analysis, we must ascertain first of
all if:
\begin{itemize}
\item the sub-process with gluon exchange in the crossed channel,
\emph{i.e.} the leading order of the BFKL resummation, dominates
over all the other sub-processes;
\item the acceptance cuts of the experiment under consideration allow us
to reach the kinematic region of the high-energy limit, where the
approximations needed for a BFKL analysis are valid.
\end{itemize}

The goal of this paper is to analyse whether the two conditions above
are fulfilled in the $\gamma^* \gamma^* \to$ hadrons process at LEP2.
Namely, we consider
\beq
\gamma^* +\gamma^*\;\longrightarrow\; {\rm hadrons},
\label{processgg}
\eeq
in $\epem$ collisions
at photon virtualities $q_i^2=-Q_i^2 < 0$, and for large
centre-of-mass energies squared $W^2=(q_1+q_2)^2$, with $q_i$ being the
momenta of the photons.
In practice we can realise the scattering (\ref{processgg}) in the process
\beq
e^+ +e^-\,\longrightarrow\, e^+ +e^- + {\rm hadrons},
\label{fullproc}
\eeq
of which \eqn{processgg} constitutes a subset.
Other contributions to the process in Eq.~(\ref{fullproc}) are, for example,
those in which the incoming $\epem$ pair annihilates into a photon or a $Z$
boson, eventually producing the hadrons and a lepton pair, or those in which
one (or both) of the two photons is replaced
by a $Z$ boson. However, it is not difficult to devise a set of cuts such
that the multiperipheral process
\beq
\begin{array}{rcl}
e^+ + e^- & \longrightarrow & e^+ + e^- + \underbrace{\gamma^* + \gamma^*} \\
 &  & \phantom{e^+ + e^- + \gamma^*\:}\bentarrow {\rm hadrons} ,
\end{array}
\label{processee}
\eeq
gives the only non-negligible
contribution to the process in Eq.~(\ref{fullproc}). One can tag
both of the outgoing leptons, and retain only those events (thus
termed {\em double-tag events}) in which the scattering angles
of the leptons are small: in such a way, the contamination due
to annihilation processes is safely negligible.
Furthermore, small-angle tagging also guarantees that the photon
virtualities are never too large (at LEP2, one typically measures
$Q_i^2\approx 10$~GeV$^2$); therefore, the contributions from processes
in which a photon is replaced by a $Z$ boson are also negligible.
Thus, it is not difficult to extract the cross section of the
process \mbox{$\gamma^*\gamma^* \to$~hadrons} from the data relevant
to the process in Eq.~(\ref{fullproc}). Double-tag events have in
fact been studied by the CERN L3 and OPAL Collaborations, at various $\epem$
centre-of-mass energies ($\sqrt{s} =$ 91 and 183 GeV~\cite{Acciarri:1999ix},
and 189-202 GeV~\cite{Abbiendi:2001tv,Achard:2001kr}).

The process (\ref{processgg}) has been analysed at
leading order~\cite{Vermaseren:1983cz} and at next-to-leading
order~\cite{Cacciari:2001cb} (NLO) in $\as$.
However, at the high end of the $W$
spectrum, the NLO prediction does not suffice to describe the data.
In this paper, we consider the part of the next-to-next-to-leading order
(NNLO) corrections which yields the dominant contribution to the total
cross section in the large-$W$ region. As we shall argue below, that
contribution comes from four quark final states. Since four parton
final states do not yield {\it per se} a finite contribution to the total
cross section, we consider a subset of them, those with gluon exchange
in the crossed channel, which are finite, and argue that they yield the
most important contribution to the total cross section in the large-$W$
region.

The paper is organised as follows: in \sec{sec:two} we set the
theoretical framework; in \sec{sec:4qamp} we consider the
four-quark contribution to the $\ord(\as^2)$ corrections, exactly
and in the high-energy limit; analysing various theoretical predictions
for rapidity and $W$ distributions in \sec{sec:rapid}, we substantiate
our claim that the most important contribution to the total cross
section in the large-$W$ region comes from four-quark production with
gluon exchange in the crossed channel, and also show that the
high-energy limit is not sufficiently accurate at LEP2 energies;
in \sec{sec:pheno} we present our
phenomenological results, by comparing our predictions to the L3
data~\cite{Achard:2001kr} (we shall not perform a comparison with
the OPAL data, which have much poorer statistics);
in \sec{sec:conc} we draw our conclusions.

\section{The theoretical framework}
\label{sec:two}

At leading order, the multiperipheral process (\ref{processee}) is
modelled by the partonic subprocess $\gamma^*\gamma^* \to q \qb$,
depicted in \fig{fig:diagr}(a),
which has been computed for massless and massive final-state
fermions~\cite{Vermaseren:1983cz}. In Ref.~\cite{Cacciari:2001cb}
the $\ord(\as)$ QCD corrections to the process (\ref{processee})
were computed for final-state massless quarks (sample diagrams are given
in \fig{fig:diagr}(b)-(c)).
\EPSFIGURE[t]{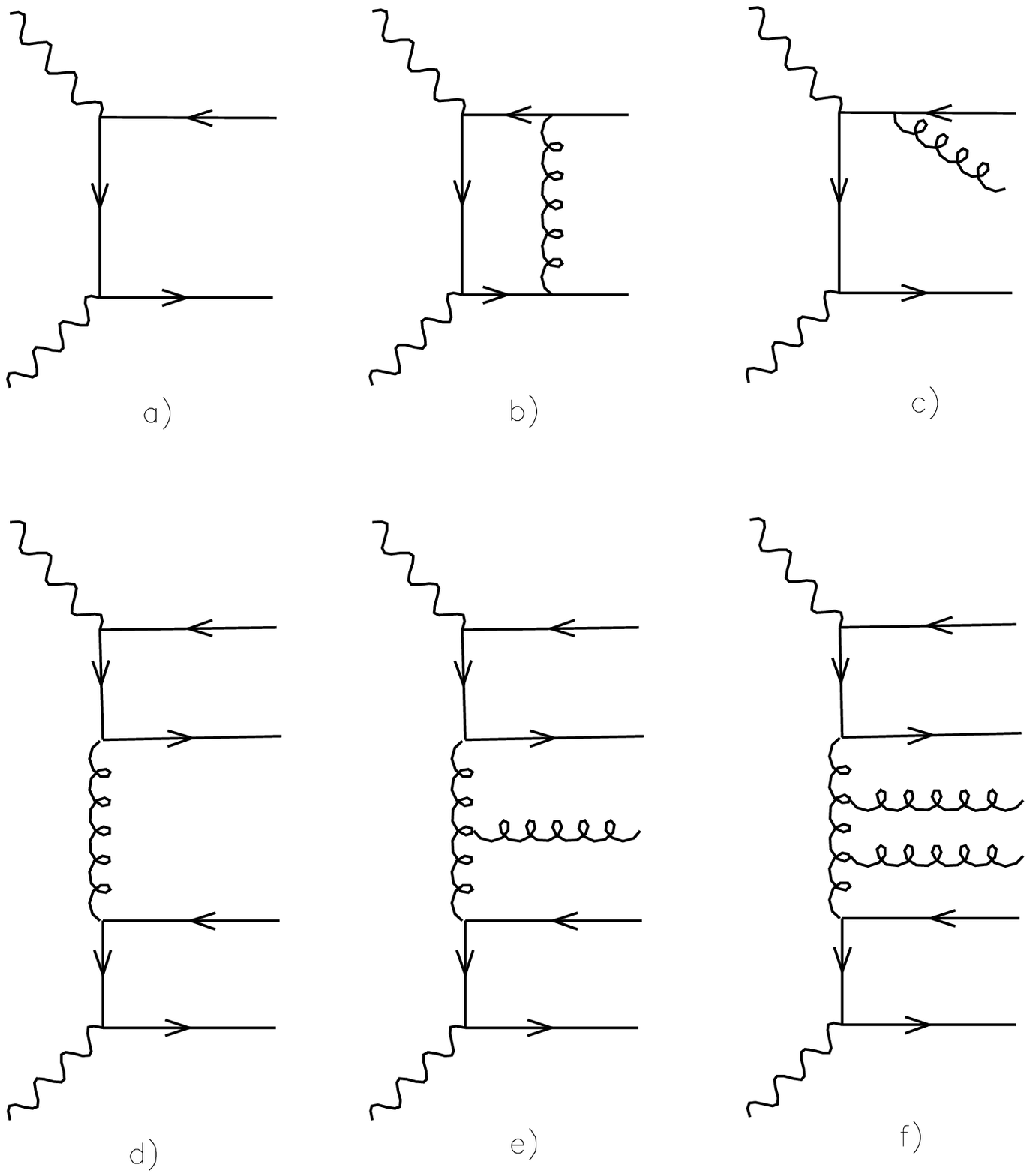,width=0.6\textwidth,clip=}
{Sample of diagrams contributing to the production of hadrons
in the collision of two off-shell photons.
\label{fig:diagr} }
However, the NLO analysis of
Ref.~\cite{Cacciari:2001cb} yields a cross section behaving as
\beq
\sigma_{\gamma^*\gamma^*}\,\sim\,1/W^2,
\label{xsec}
\eeq
modulo logarithmic corrections. Thus, it is only propaedeutic to the
BFKL resummation, whose leading-order term is based on
the exchange of a gluon in the crossed channel, which appears only at
$\ord(\as^2)$ (\fig{fig:diagr}(d)). The BFKL resummation then builds up
gluon emission along the gluon exchanged in the crossed channel (the
first rungs of the ladder are represented in \fig{fig:diagr}(e)-(f)).
The diagrams represented by \fig{fig:diagr}(d) are expected to yield
a cross section which, away from the threshold and the kinematic limit,
is weakly dependent on $W$. The additional
gluon emissions build up the logarithmic corrections which the
BFKL theory resums, so that the full cross section is expected to
behave as
\beq
\sigma_{\gamma^*\gamma^*}\,\sim\, \sum_{j=0}^\infty a_{0j} \as^j\
+\ a_1 \as^2 \sum_{j=0}^\infty (\as L)^j\ +\ a_2 \as^2 \sum_{j=0}^\infty
\as(\as L)^j\ + \cdots,
\label{BFKLxsec}
\eeq
where \mbox{$L=\log(W^2/\mu_{\sss\rm W}^2)$} is a large logarithm,
and the quantity $\mu_{\sss\rm W}^2$ is a
mass scale squared, typically of the order of the crossed-channel momentum
transfer and/or of the photon virtualities.
In \eqn{BFKLxsec}, the second and third sums collect
the contributions which feature \emph{only} gluon exchange in the crossed
channel, the second (third) sum resumming the BFKL (next-to-)leading
logarithmic corrections; the $a_1$, $a_2$ coefficients behave
like $1/\mu_{\sss\rm W}^2$. The ellipses refer to logarithmic corrections
beyond the next-to-leading accuracy. The first sum in \eqn{BFKLxsec}
is a fixed-order expansion in $\as$ starting at $\ord(\as^0)$, and
collects the contributions which do not feature gluon exchange in the
crossed channel; the $a_{0j}$ coefficients behave like
$1/W^2$\footnote{The $a_{02}$ coefficient may
feature terms which behave like $1/(W \mu_{\sss\rm W})$ and
arise from the interference between diagrams with gluon exchange
in the crossed channel and diagrams with quark exchange in the
crossed channel. These interference terms have been analysed in
\sec{sec:rapid} (see the discussion of \vfig{fig:yshape}).}.
Thus, it is clear that the second and third sums of \eqn{BFKLxsec}
will eventually dominate over the first sum in the
asymptotic energy region $W\to\infty$. The second sum of \eqn{BFKLxsec}
has been analysed in the region $W^2\gg\mu_{\sss\rm W}^2$, by computing
in the high-energy limit the $a_1$ coefficient in the
massless~\cite{Brodsky:1997sd,Bartels:1996ke} and in the
massive~\cite{Bartels:2000sk} case. As mentioned above,
the $a_{00}$ term in the first sum has been computed in
Ref.~\cite{Vermaseren:1983cz} for massless and massive final-state quarks,
while the $a_{01}$ term has been computed in Ref.~\cite{Cacciari:2001cb},
for massless final-state quarks. In the next paragraph, we shall illustrate
that at present a calculation of the $a_{02}$ term is unfeasible.
In this work, we compute \emph{exactly} the $a_1$ coefficient in the
massless limit, and add it
to the $a_{00}$ and  $a_{01}$ terms. However, we do not perform the
resummation, \emph{i.e.} we consider only the $j=0$ term in
the second sum of \eqn{BFKLxsec}.

\EPSFIGURE[t]{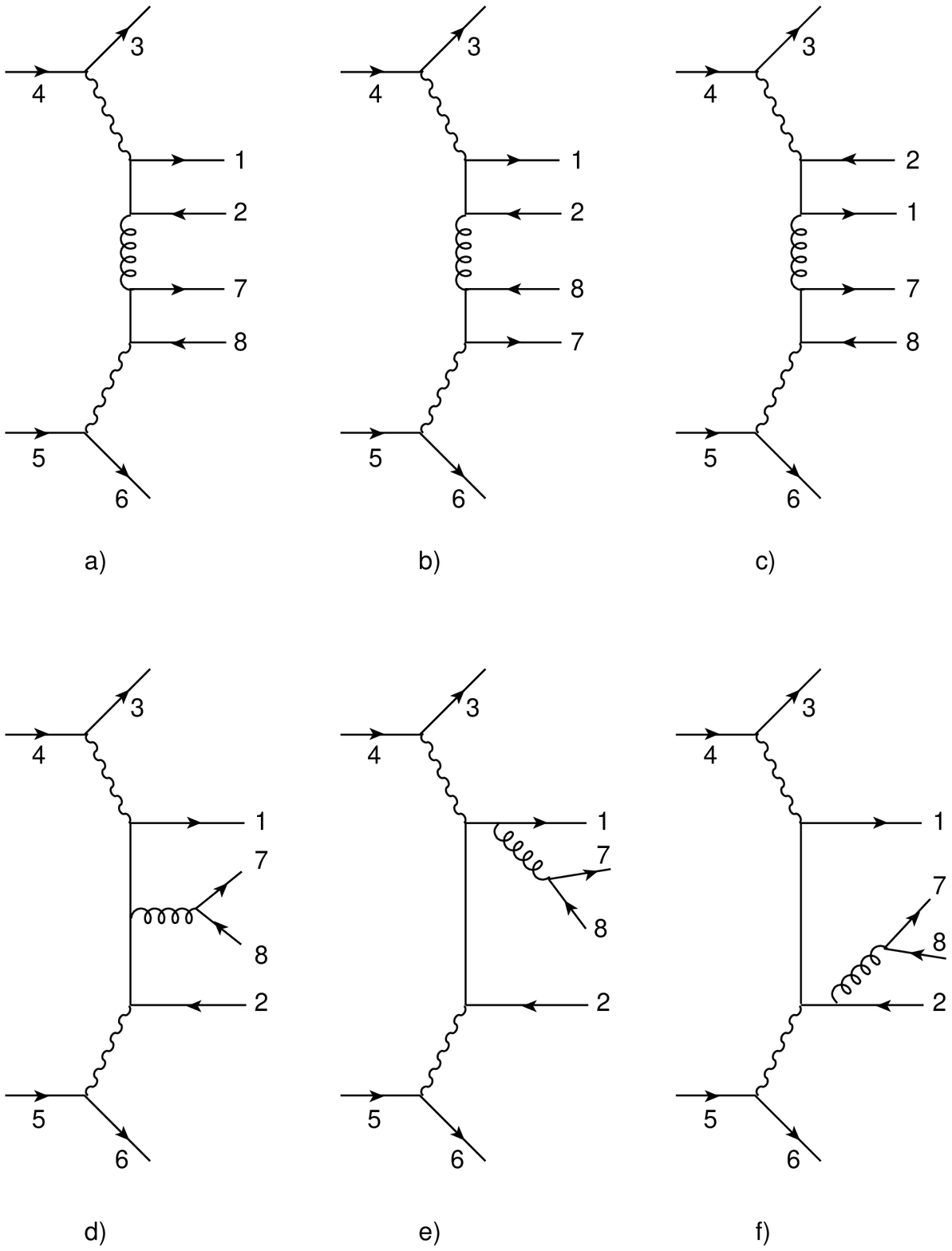,width=10cm}
{Amplitude for $e^+e^- \to e^+e^-\gamma^*\gamma^* \to e^+e^-
q\bar q Q\bar Q$ scattering. Figures $(a)$, $(b)$ and $(c)$ correspond
to the functions $g_a$, $g_b$ and $g_c$ in \eqn{a8treesym}; figures
$(d)$, $(e)$ and $(f)$ contribute to the function $f$.
\label{fig:gammastarqqqq}  }

In examining the radiative corrections to
the process (\ref{processee}), we first note that at present a
calculation of the full $\ord(\as^2)$ corrections to the
total and to the inclusive jet and dijet cross sections
is unfeasible, since it would require a computation of two-loop
amplitudes, including double box diagrams with two off-shell legs which are
not known. A calculation of the $\ord(\as^2)$ corrections to
the inclusive three-jet rate is feasible but difficult,
and given the very limited experimental statistics available
it would have at present only an academic value.
Therefore, we shall limit ourselves to the analysis of
the contribution of the four parton final states to the $\ord(\as^2)$
corrections. In particular, the final states can be made of either
two quark pairs or a quark pair and two gluons. In \fig{fig:gammastarqqqq}
the diagrams with two final-state quark pairs are represented.

The treatment of the four parton final states poses some
additional problems, because as far as the total, the inclusive jet and
the dijet cross sections are concerned, only the contribution of the
four-quark diagrams with gluon exchange in the crossed channel
(\fig{fig:gammastarqqqq}(a)-(c)) is infrared finite.
Henceforth, we shall term the diagrams with gluon exchange
in the crossed channel the {\it $g$ class}. The diagrams of the $g$
class are by themselves gauge invariant and as described in the former
paragraph, they are expected to yield a cross section
which is only logarithmically dependent on $W^2$. Thus, as $W^2$ grows,
they are expected to dominate over the diagrams with quark exchange
in the crossed channel.
The diagrams with two final-state quark pairs and quark
exchange in the crossed channel (\fig{fig:gammastarqqqq}(d)-(f)) and
the diagrams with a quark pair and two gluons in the final state, present
infrared divergences when one or both the final-state gluons become collinear
to the parent quarks or between themselves, when two or three final-state
quarks become collinear, or when the gluon emitting the quark
pair becomes soft. We shall term the diagrams with a quark pair and two gluons
in the final state, and the diagrams with two final-state quark pairs
and quark exchange in the crossed channel the {\it $f$ class}.
The diagrams of the $f$ class would all be part
of the proper NNLO corrections to the process
$\gamma^*\gamma^* \to q \qb$, and  after cancellation of their
infrared divergences by the corresponding virtual terms their
contribution to the cross section is expected to have a dependence on
$W^2$ like in \eqn{xsec}, up to logarithmic corrections.
They would contribute the $a_{02}$ coefficient in \eqn{BFKLxsec}.

\section{The four-quark contribution to the $\boldsymbol{\ord(\as^2)}$
corrections}
\label{sec:4qamp}

When calculating
the four-quark contribution to $\gamma^* \gamma^* \to$ hadrons,
\beq
e^+e^- \to e^+e^-\gamma^*\gamma^* \to e^+e^- q\bar q Q\bar Q\, ,
\eeq
we assume that all the produced quarks are massless
and that the virtualities $Q_{1,2}^2$ are low
enough that contributions of virtual $W,Z$ bosons can be neglected.
Indeed, for the four-quark contributions at 200\,GeV centre-of-mass
energy $\langle W \rangle \simeq$ 40\,GeV and
$\langle Q_i^2 \rangle \simeq 13\,\mbox{GeV}^2$.
We have re-computed the tree amplitudes for four-quark
production~\cite{Barger:1990yd} using the spinor products of
\app{sec:spipros}. The contributing diagrams are shown in
\fig{fig:gammastarqqqq}, featuring gluon (\fig{fig:gammastarqqqq}$(a)-(c)$)
or quark (\fig{fig:gammastarqqqq}$(d)-(f)$) exchange in the crossed channel.
For two quark pairs of different flavour, we have
\beq
\cA_8(1_q,2_\qb;3_\ell,4_\ellb,5_{\ellb'},6_{\ell'};
7_Q,8_\Qb)  = 4 e^4 \gs^2\ T^a_{i_1\ib_2} T^a_{i_7\ib_8}\
A_8(1,2;3,4,5,6;7,8) \, ,\label{A8}
\eeq
with $\{1,2\}$ and $\{7,8\}$ the quark pairs, and $\{3,4\}$ and $\{5,6\}$
the lepton pairs, and where the colour-stripped sub-amplitude $A_8$
depends on the momenta and helicities of the external particles.
By convention, all particles
are taken as outgoing, thus an incoming fermion of a given helicity
is represented by an outgoing antifermion of the opposite helicity.
$A_8$ can be divided into the functions $a_8$, $b_8$ and $c_8$,
\bea
\lefteqn{ A_8(1_q,2_\qb;3_\ell,4_\ellb,5_{\ellb'},6_{\ell'};7_Q,8_\Qb)
= Q_{f_q} Q_{f_Q} a_8(1,2;3,4,5,6;7,8) } \nonumber\\
&& \qquad + Q_{f_q}^2 b_8(1,2;3,4,5,6;7,8) +  Q_{f_Q}^2 c_8(1,2;3,4,5,6;7,8)
\, ,\label{a8treesym}
\eea
with $Q_{f_{q(Q)}}$ the electric charge fraction of the quark $q(Q)$ of flavour
$f_{q(Q)}$. The calculation of the functions $a_8$, $b_8$ and $c_8$ as
well as the one of the related production rate is detailed in
\app{sec:helamp}.

\subsection{The high-energy limit}

In \sec{sec:two}, we claimed that when the squared hadronic energy
$W^2$ is much larger than the typical momentum transfer $\mu^2_{\sss W}$,
and radiative corrections to the gluon exchanged between two quark pairs
are considered, the large logarithms of type $\ln(W^2/\mu^2_{\sss W})$
which ensue
can be resummed through the BFKL ladder. In fact, the resummation of the
BFKL ladder requires more restrictive kinematics, where the rapidities
of quarks which do not belong to the same quark pair are strongly ordered,
\beq
\eta_1 \simeq \eta_2 \gg \eta_7 \simeq \eta_8 \qquad {\mbox or} \qquad
\eta_1 \simeq \eta_2 \ll \eta_7 \simeq \eta_8 \, .\label{hekinem}
\eeq
\eqn{hekinem} defines the
\emph{high-energy limit} for $\gamma^*\gamma^* \to$ hadrons.
When the strong rapidity ordering (\ref{hekinem}) occurs, the
diagrams with gluon exchange in the crossed channel yield the
dominant contribution\footnote{Note that the reverse is not true. In fact
the diagrams with gluon exchange in the crossed channel may dominate over
the diagrams with quark exchange
well before the high-energy limit is realised. This issue is
discussed in \sec{sec:rapid}.}, and in the
amplitude (\ref{a8treesym}) the functions $b_8$ and $c_8$ can be neglected.
Thus the two final-state quark pairs can be treated as non-interacting, and
in the squared amplitude we can take the quark flavours as always distinct,
\beq
\frac12 \sum_{f_q,f_Q} |A_8(1,2;3,4,5,6;7,8)|^2 =
\frac12 ( Q_u^2 n_u + Q_d^2 n_d )^2
|a_8(1,2;3,4,5,6;7,8)|^2\, ,\label{sums_distinct}
\eeq
with $Q_u=2/3$, $Q_d=-1/3$ and $n_{u(d)}$ the number of up(down)-type
quarks. In \eqn{sums_distinct} the factor of $1/2$ appears in
order to avoid double counting, since the amplitudes are symmetric with
respect to the interchange of the two quark lines.
Thus for a fixed lepton-helicity configuration,
e.g~$(3_\ell^-,4_\ellb^+,5_{\ellb'}^+,6_{\ell'}^-)$, the production rate is
\bea
\lefteqn{ \d\sigma(3_\ell^-,4_\ellb^+,5_{\ellb'}^+,6_{\ell'}^-) }
\label{LOcross_distinct}\\ &=&
{1\over 2s} \d\cP_6  16 \left(N_c^2 -1\right) (4\pi\aem )^4 (4\pi\as)^2
\frac12 ( Q_u^2 n_u + Q_d^2 n_d )^2
\nonumber\\ &&\times
\Bigl[ |a_8(1^-,2^+;3^-,4^+,5^+,6^-;7^-,8^+)|^2
+ |a_8(2^-,1^+;3^-,4^+,5^+,6^-;7^-,8^+)|^2 \nonumber\\ && \
+\, |a_8(1^-,2^+;3^-,4^+,5^+,6^-;8^-,7^+)|^2 +
|a_8(2^-,1^+;3^-,4^+,5^+,6^-;8^-,7^+)|^2 \Bigr]\, ,\nonumber
\eea
where the six-particle phase space is given in \eqn{6phase}.
The other lepton-helicity configurations are simply obtained by
exchanging the labels 3 and 4 and/or 5 and 6 in \eqn{LOcross_distinct}
(see \app{sec:helamp}). The unpolarised rate is given by averaging over
the rates for the four lepton-helicity configurations.

Since in the high-energy limit the two final-state quark pairs behave
effectively as if they were two independent scattering centres, the
amplitude (\ref{a8treesym}) with the functions $b_8$ and $c_8$ set
to zero is expected to factorise into
two high-energy coefficient functions, usually termed impact factors,
for the process $e  g^*\to e q \bar{q}$, where $g^*$ is the off-shell
gluon which is exchanged in the crossed channel.
In the high-energy limit, the amplitude (\ref{a8treesym})
can then be used to derive such impact factors. However, it is easier to
invoke high-energy factorisation and to
derive the impact factor for $e  g^*\to e q \bar{q}$ from a simpler process,
e.g. from the scattering amplitudes $e g \to q \qb g$ or $e Q \to q \qb Q$
which appear typically in DIS processes. Then one can use two such impact
factors, one in the forward and one in the backward kinematics,
connected by a gluon exchanged in the crossed channel, in order
to obtain the amplitude for $\epem \to q \qb Q \bar{Q}$
in the high-energy limit. We denote the impact factor for the
$e  g^*\to e q \bar{q}$ process in the forward (backward) kinematics,
evaluated in the $\gamma^*\gamma^*$
centre-of-mass frame, as $V_{f(b)}(p_\ellb;p_\ell,p_q,p_\qb)$ (remember
that all momenta are outgoing, hence the dependence on momenta of two
particles and two antiparticles),
and derive it in \app{sec:dif} (see \eqn{disifpp2}).
Then the amplitude (\ref{A8}) factorises as
\bea
\lefteqn{ \cA_8(1_q,2_\qb;3_\ell,4_\ellb,5_{\ellb'},6_{\ell'};
7_Q,8_\Qb) } \nonumber\\ && \hspace{-0.2cm} = 2\,\sgg \left[ \gs e^2 Q_{f_q}\,
T^c_{i_1\ib_2}\,\sqrt{2} V_f(4_\ellb;3_\ell,1_q,2_\qb) \right] {1\over t}
\left[ \gs e^2 Q_{f_Q}\, T^c_{i_7\ib_8}\,\sqrt{2} V_b(6_{\ell'};5_{\ellb'},7_Q,8_\Qb)
\right]\, ,\label{gammafac}
\eea
with $t = q^2$, where $q = - \sum_{i=1}^4 p_i$ is the momentum transfer.
The two impact factors for $e g^*\to e q \qb$ can be extracted through
re-labelling from \eqnss{disifpp2}{backward}.
Each of those impact factors can be decomposed further into a lepton
current and an impact factor for $\gamma^* g^*\to q \qb$ (see
\eqnss{ifgamma}{Vgamma-qqb} ).

In computing the square of the amplitude, we must sum over helicity, colour
and flavour of the quarks, however in this case the flavour sum is trivial
since the two impact factors do not interfere and we can treat the quark
flavours as always distinct. The production rate is
\bea
\d\sigma(3_\ell^-,4_\ellb^+,5_{\ellb'}^+,6_{\ell'}^-) &=&
{1\over 2s} d\cP_6\, 16 \left(N_c^2 -1\right)
(4\pi\aem )^4 (4\pi\as)^2
(Q_u^2 n_u + Q_d^2 n_d)^2  \nonumber\\ &\times&
{\sgg^2\over t^2}
\Big[  \left( |V_f(4_\ellb^+;3_\ell^-,1_q^-,2_\qb^+)|^2 +
|V_f(4_\ellb^+;3_\ell^-,1_q^+,2_\qb^-)|^2 \right)
\nonumber\\ && \qquad\; \times
\left( |V_b(5_{\ellb'}^+;6_{\ell'}^-,7_Q^-,8_\Qb^+)|^2 +
|V_b(5_{\ellb'}^+;6_{\ell'}^-,7_Q^+,8_\Qb^-)|^2 \right) \Big]\,,
\label{eq:HElimit}
\eea
which constitutes the high-energy factorisation of \eqn{LOcross_distinct}.
Each of the two rapidity orderings of \eqn{hekinem} yield the same contribution
to \eqn{eq:HElimit}. Thus we have included them by taking only
the first of the two and deleting the double counting factor 1/2.
As in Sect. 2.3 of Ref.~\cite{Cacciari:2001cb},
the phase space (\ref{6phase}) for the $e^+e^- q\bar q Q\bar Q$ final state
can be factorised into hadronic and leptonic phase spaces,
\beq
\d{\cal P}_6 = \d\Gamma(p_3,p_5)\, \d{\cal P}_4(p_1,p_2,p_7,p_8;k_1+k_2)\, ,
\eeq
with $k_1 = p_4 - p_3$ and $k_2 = p_6 - p_5$ the momenta of the
virtual photons (here we have inverted the direction of $p_4$ and $p_6$
in order to have them incoming), and
\bea
\d\Gamma &=& {d^3 p_3\over (2\pi)^3 2p^0_3} {d^3 p_5\over (2\pi)^3 2p^0_5}
\nonumber\\ \d{\cal P}_4 &=& \prod_{i=1,2,7,8} {d^3 p_i\over (2\pi)^3 2p^0_i}\,
(2\pi)^4 \, \delta^4(k_1 + k_2 - p_1 - p_2 - p_7 - p_8)\,
,\label{lephadphase}
\eea
the leptonic and hadronic phase spaces, respectively.

In the high-energy limit, momentum conservation for
$e^+e^- \to e^+e^-\gamma^*\gamma^* \to e^+e^- q\bar q Q\bar Q$
implies that, in the $\gamma^*\gamma^*$ centre-of-mass
frame\footnote{\eqn{gammakin}
is valid also in $e^+e^-$ centre-of-mass frame by adding
$p_{3\perp} + p_{5\perp}$ to the right hand side of the third line.},
\bea
k_1^+ &\simeq& p_1^+ + p_2^+ \nonumber\\
k_2^- &\simeq& p_7^- + p_8^- \label{gammakin}\\
0 &=& p_{1\perp} + p_{2\perp} + p_{7\perp} + p_{8\perp} \nonumber
\eea
where we use light-cone coordinates: $p^\pm = p^0 \pm p^z$ and for the
two-dimensional vector $p_\perp$ complex transverse coordinates
$p_{\perp} = p^x + i p^y$ (see \app{sec:dif}).
Momentum conservation (\ref{gammakin}) allows us to factorise the hadronic
phase space (\ref{lephadphase}) further,
\bea
\d{\cal P}_4 &=& 2 \left( \prod_{i=1,2} {d^3 p_i\over (2\pi)^3 2p^0_i}\,
2\pi \, \delta(k_1^+ - p_1^+ - p_2^+) \right)\, \left( \prod_{i=7,8}
{d^3 p_i\over (2\pi)^3 2p^0_i}\, 2\pi \, \delta(k_2^- - p_7^- - p_8^-) \right)
\nonumber\\ &&\times (2\pi)^2 \, \delta^2(p_{1\perp}
+ p_{2\perp} + p_{7\perp} + p_{8\perp})\, .\label{heps}
\eea
The terms in round brackets in the first line are the phase spaces
for the two impact factors. They are
connected by transverse momentum conservation only. The overall
factor of 2 in $\d{\cal P}_4$ comes from the Jacobian of the light-cone
coordinates. \eqn{heps} can be immediately generalised to the emission of
a BFKL gluon ladder between the impact factors.

Fixing
\beq
x_a = {p_1^+\over p^+_1 + p_2^+} = 1 - {\tilde x}_a\:,\qquad
x_b = {p_7^-\over p^-_7 + p_8^-} = 1 - {\tilde x}_b\:,
\eeq
the phase space (\ref{heps}) can be re-written as
\bea
\d{\cal P}_4 &=& {1\over (4\pi)^2} {1\over 2k_1^+ k_2^-}
\left( {dx_a\over x_a(1-x_a)} {d^2p_{1\perp} \over (2\pi)^2} \right)
\left( {dx_b\over x_b(1-x_b)} {d^2p_{7\perp} \over (2\pi)^2} \right) \nn\\
&& \times
{d^2q_{a\perp} \over (2\pi)^2} {d^2q_{b\perp} \over (2\pi)^2}
(2\pi)^2 \, \delta^2(q_{a\perp} - q_{b\perp})
\label{eq:HEPS}
\eea
with $q_a= k_1 - p_1 - p_2$ and $q_b = p_7 + p_8 - k_2$. Note that in
the $\gamma^*\gamma^*$ centre-of-mass frame, the momenta of the
virtual photons are (in the light-cone notation of \app{sec:dif})
$k_1=(k_1^+,k_1^-;0_\perp)$ and $k_2=(k_2^+,k_2^-;0_\perp)$, with
virtualities $k_1^2 = k_1^+ k_1^- = - Q_1^2$ and
$k_2^2 = k_2^+ k_2^- = - Q_2^2$. In the high-energy limit, $k_1^+\gg k_1^-$
and $k_2^-\gg k_2^+$, thus the centre-of-mass energy is
$s_{\gamma^*\gamma^*} = (k_1 + k_2)^2 \approx k_1^+ k_2^-$.

\section{Theoretical predictions}
\label{sec:rapid}

In this section we present the results obtained by considering the
contribution of the four
parton final state to the cross section for $\gamma^* \gamma^* \to$
hadrons. The four parton final state is $\ord(\aem^4\as^2)$, however
from the stand point of both the electromagnetic and the strong corrections
it is a leading order calculation, thus the dependence of either the
electromagnetic or the strong coupling on the respective scales is maximal.

As far as $\aem$ is concerned, we have chosen to set the scales on
an event-by-event basis to the virtualities of
the exchanged photons; hence, we replace the Thomson value $\alpha_0
\simeq 1/137$ by $\aem(Q_i^2)$, as in Ref.~\cite{Cacciari:2001cb}.
This choice better describes the
effective strength at which the electromagnetic interaction takes place.
In addition, we treat independently the two photon legs: thus, in the
formul\ae\ relevant to the cross sections, $\alpha_{\rm em}^4$ has to be
understood as \mbox{$\alpha_{\rm em}^2(Q_1^2)\alpha_{\rm em}^2(Q_2^2)$}.

As far as $\as$ is concerned, we define a default scale $\mu_0$ so
as to match the order of magnitude of the inverse of the interaction
range~\cite{Cacciari:2001cb},
\beq
\mu_0^2 = \frac{Q_1^2+Q_2^2}{2} +
\left(\frac{p_{1\perp} + p_{2\perp} + p_{7\perp} + p_{8\perp}}
{2}\right)^2 \: .\label{defscale}
\eeq
Scale choices other than (\ref{defscale}) have been considered in
Ref.~\cite{Cacciari:2001cb}.
The renormalisation scale $\mu$ entering $\as$ is set
equal to $\mu_0$ as a default value, and equal to $\mu_0/2$ or $2\mu_0$ when
studying the scale dependence of the cross section. In Eq.~(\ref{defscale}),
the $p_{i\perp}$ are the transverse energies of the outgoing partons.
Since the hard process is
initiated by the two virtual photons, the proper frame to study its
properties is the $\gamma^*\gamma^*$ centre-of-mass one. Therefore,
when talking about transverse energies, whether in a total
or in a jet cross section, this frame will be always understood.

We evolve $\as$ to next-to-leading log accuracy,
with $\as(M_{\sss Z})=0.1181$~\cite{Groom:2000in} (in $\overline{{\rm MS}}$
at two loops and with five flavours, this implies
$\Lambda_{\overline{{\rm MS}}}^{(5)} = 0.2275$ GeV).
The choice of the two-loop running is due to the fact that
we are going to use a full NLO calculation augmented by a partial
$\ord(\as^2)$ contribution (the diagrams of the $g$ class only).
When presenting numerical results we use five massless flavours in the
cross section formulae (\ref{eq:HElimit}) and (\ref{sums}).

In exploring the footprints of the BFKL resummation in $\epem$ collisions, it
is customary to introduce the variable $Y$,
\beq
Y=\log\frac{y_1 y_2 s}{\sqrt{Q_1^2 Q_2^2}}\, ,\label{YD}
\eeq
where the variables $y_i$ are proportional to the light-cone momentum fraction
of the virtual photons,
\beq
y_i = {q_i^0 + q_i^3 \over \sqrt{s}} = 1 - {2E_i\over \sqrt{s}}
\cos^2{\theta_i\over 2}\, ,\qquad i=1,2\, ,\label{yi}
\eeq
where $E_i$ and $\theta_i$ are the energies and scattering angles of the
outgoing electron and positron in the $e^+e^-$ centre-of-mass frame.
For large $Y$, we have $y_1 y_2 s\approx W^2$, {\it i.e.} the
$Y$ variable parametrises the ratio of the hadronic energy over a typical
momentum transfer, thus it is a variable which is suitable for analyses
of the BFKL type.

\EPSFIGURE[t]{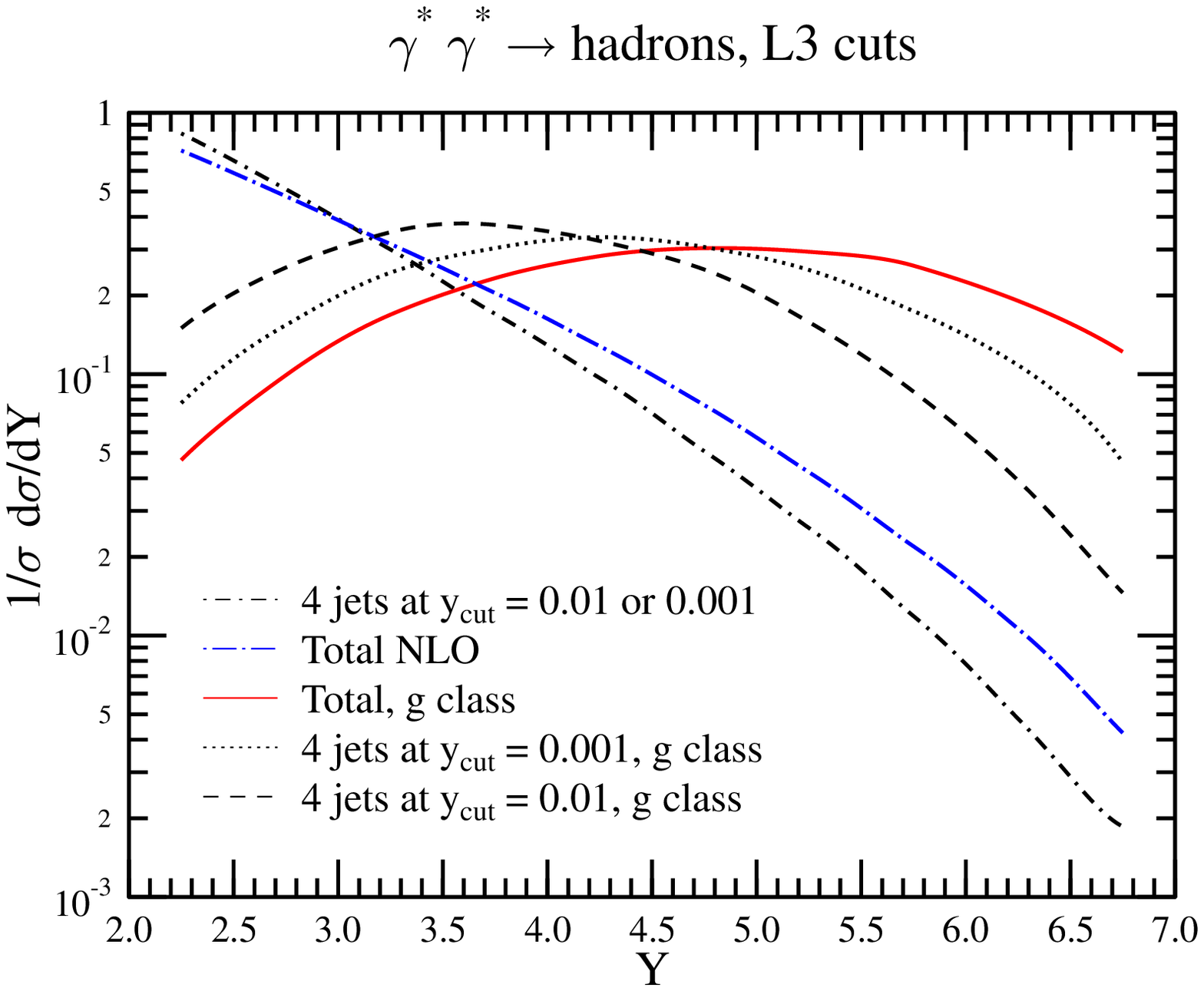,width=14cm}
{$Y$ distribution in different production
rates. Namely, in the total NLO cross section (dot-long-dashed line),
in the total cross section due to diagrams of the $g$ class only
(solid line), in the four-jet cross section (dot-short-dashed line) and
in the four-jet cross section with diagrams of the $g$ class only at
$y_{\rm cut} = 0.01$ (dashed line) and at $y_{\rm cut} = 0.001$ (dotted
line). The relative normalisation of the curves has been rescaled, in such a
way that the area under each of the curves is the same.
\label{fig:yshape}  }

In \vfig{fig:yshape}, we plot the $Y$ distribution for different production
rates. Namely, for the total NLO cross section (dot-long-dashed line),
for the total cross section due to diagrams of the $g$ class only
(solid line), for the four-jet cross section at $y_{\rm cut} = 0.01$ or
0.001 (dot-short-dashed line) and
for the four-jet cross section with diagrams of the $g$ class only at
$y_{\rm cut} = 0.01$ (dashed line) and at $y_{\rm cut} = 0.001$ (dotted
line). Throughout this plot and henceforth,
we use the acceptance cuts of the CERN L3 Collaboration~\cite{Achard:2001kr},
namely, the lepton energies $E_i$ are larger than 40\,GeV, the lepton tagging
angles $\theta_i$ are between 30 and 66\,mrad and the hadronic energy $W$
is larger than 5\,GeV. As far as the photon virtualities are concerned,
the cuts above imply that $Q_{1,2}^2\gtrsim 4\ \mbox{GeV}^2$.
For the cross section due to diagrams of the $g$ class only,
we use \eqn{LOcross_distinct}. For the four-jet cross section, the
four quark final states have been computed through the formul\ae\ of
\app{sec:helamp}; the final states with two quarks and two gluons have been
generated with the help of MADGRAPH~\cite{Stelzer:1994ta}, which
has been used also to check numerically the four-quark amplitudes of
\app{sec:helamp}.
In the jet cross sections, we define the jets through a $k_{\sss T}$
algorithm~\cite{Catani:1991hj}. The jet size is set by the $y_{\rm cut}$
variable. In \fig{fig:yshape} and \ref{fig:etashape} the normalisation
of the curves is not relevant, since the contribution of the diagrams
of the $f$ class to the total cross section cannot be inferred from the
four-jet cross section, due to the lack of virtual corrections.
Thus the relative normalisation of the curves has been rescaled, in such a
way that the area under each of the curves is the same. Note that in
\fig{fig:yshape} the shape of the four-jet cross section
(dot-short-dashed line) is largely independent of the chosen $y_{\rm cut}$.
In fact, there is basically no difference between the
dot-short-dashed line at $y_{\rm cut}$ = 0.01 or 0.001.
At large $Y$ the four-jet cross section has a similar shape
as the total NLO cross section (dot-long-dashed line).
On the contrary, at large $Y$
the four-jet cross section with diagrams of the $g$ class only gets a
larger and larger contribution as $y_{\rm cut}$ goes from 0.01 (dashed line)
to 0.001 (dotted line).
In addition, the diagrams of the $g$ class are by
themselves infrared finite. Thus in the four-jet cross section
from diagrams of the $g$ class only, we can take the
limit $y_{\rm cut}\to 0$ and obtain the total cross section
(solid line). This has the most open shape at large $Y$,
which hints that at large $Y$ we should expect a substantial
contribution from the diagrams of the $g$ class to the total cross
section at $\ord(\as^2)$. The interference terms between the
diagrams of the $g$ class and those of the $f$ class, \emph{i.e.}
the terms $2[ Q_{f_q}^3 Q_{f_Q} {\rm Re}(a_8^*b_8)+Q_{f_q} Q_{f_Q}^3 {\rm
Re}(a_8^*c_8)]$ in \eqn{sums}, give also a finite contribution to the
total cross section. We have checked that they yield a curve that is
similar in shape to the total NLO cross section. Compared to the
diagrams of the $g$ class (solid line), they yield an increase of at
most 10\,\% only in the small-$Y$ region, their contribution being
negligible at large $Y$. Thus we shall neglect them henceforth.

\EPSFIGURE[t]{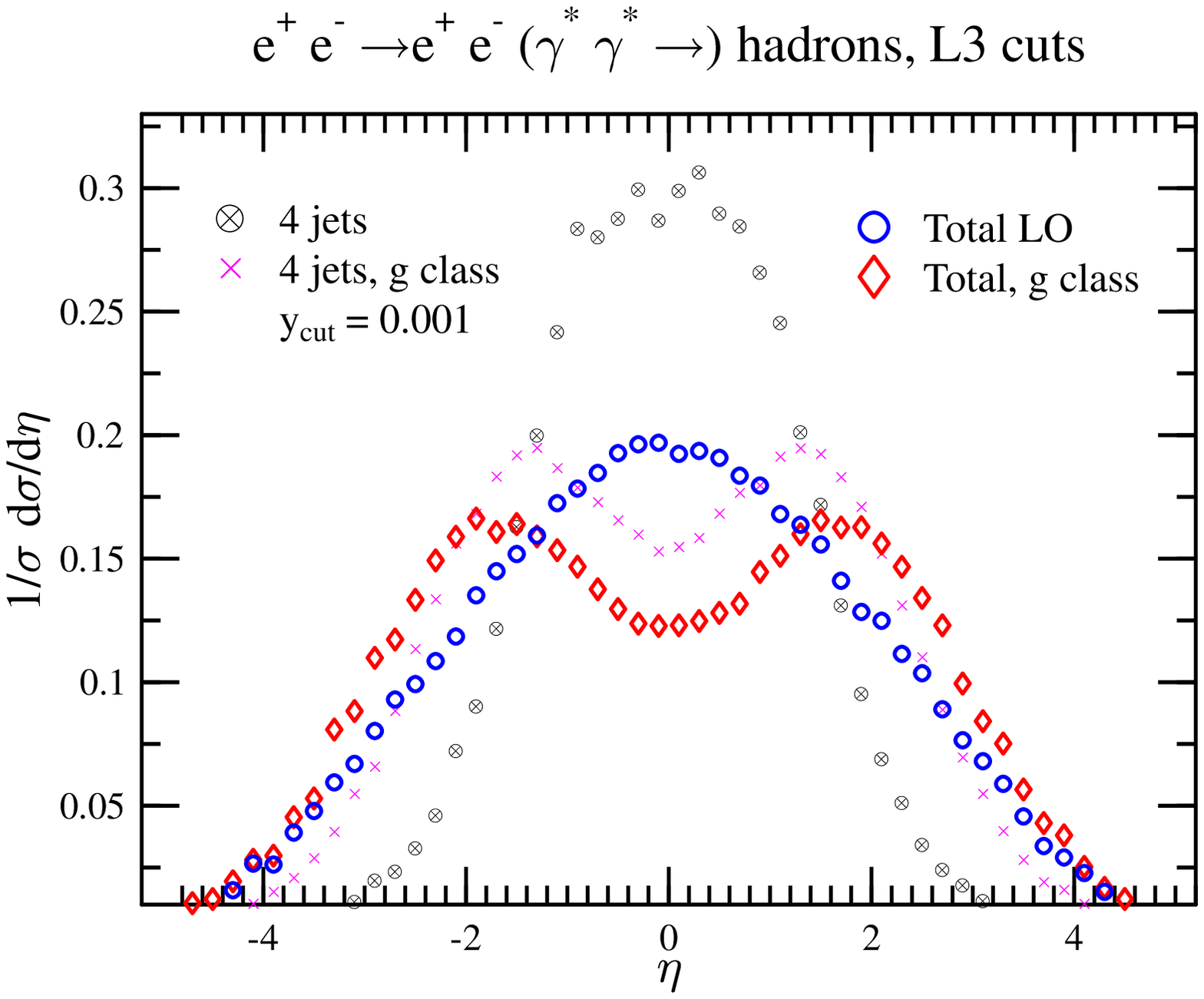,width=14cm}
{Rapidity distribution of one of the final-state partons,
in different production rates. Namely, in the total LO cross section
(\emph{circles}), in the total cross section due to diagrams of the
$g$ class only (\emph{diamonds}), in the four-jet cross section
(\emph{circles with crosses}) and in the four-jet cross section with
diagrams of the $g$ class only (\emph{crosses}).
\label{fig:etashape}  }

In \vfig{fig:etashape}, we plot the rapidity distribution of one of the
final-state partons, in
different production rates. Namely, in the total LO cross section
(\emph{circles}), in the total cross section due to diagrams of the
$g$ class only (\emph{diamonds}), in the four-jet cross section
(\emph{circles with crosses}) and in the four-jet cross section with
diagrams of the $g$
class only (\emph{crosses}).  Note that, as expected, in the total LO cross
section~\cite{Cacciari:2001cb}, which features two final-state partons
with quark exchange in the crossed channel, the partons produced
populate mainly the central rapidity region (\emph{circles}).
That is true also for the four-jet cross section (\emph{circles with crosses}).
Instead, in the four-jet cross section from diagrams of the $g$ class only
(\emph{crosses}), at $y_{\rm cut}$ = 0.001, the produced quarks populate
mainly the forward and backward rapidity regions. However, the shape of
the \emph{circles with crosses} curve depends on $y_{\rm cut}$ very mildly,
while that of the \emph{crosses} curve depends strongly on the chosen
$y_{\rm cut}$. To understand how the latter comes about, we recall
that the diagrams of the $g$ class feature two quark pairs separated by a
gluon exchanged in the crossed channel, and therefore susceptible of being
produced at large rapidity. In fact, the probability of finding one of
them in the forward and backward rapidity regions grows as $y_{\rm cut}$
becomes smaller. In the limit $y_{\rm cut}\to 0$ we obtain the
total cross section from
diagrams of the $g$ class only (\emph{diamonds}). That shows that
it is more likely to produce the quarks in the forward and backward
rapidity regions than it is to produce them in the central region\footnote{The
position of the peaks as well as the depletion in the central
rapidity region depends also on the cut on $Y$, {\it e.g.}, if we use the
cut $Y \ge 5$, the peaks move to about $\pm 3$ and become more
pronounced.}.
Finally, we recall that in a full $\ord(\as^2)$ calculation of the total
cross section, at present unfeasible, the diagrams of the $f$ class
are expected to yield a small correction to the NLO total cross section.
Thus in a full NNLO calculation of the total cross section in the large
$Y$ region, we expect the
rapidity distribution of one of the final-state partons to be roughly a
combination of the \emph{circles} and \emph{diamonds} curves.

\EPSFIGURE[t]{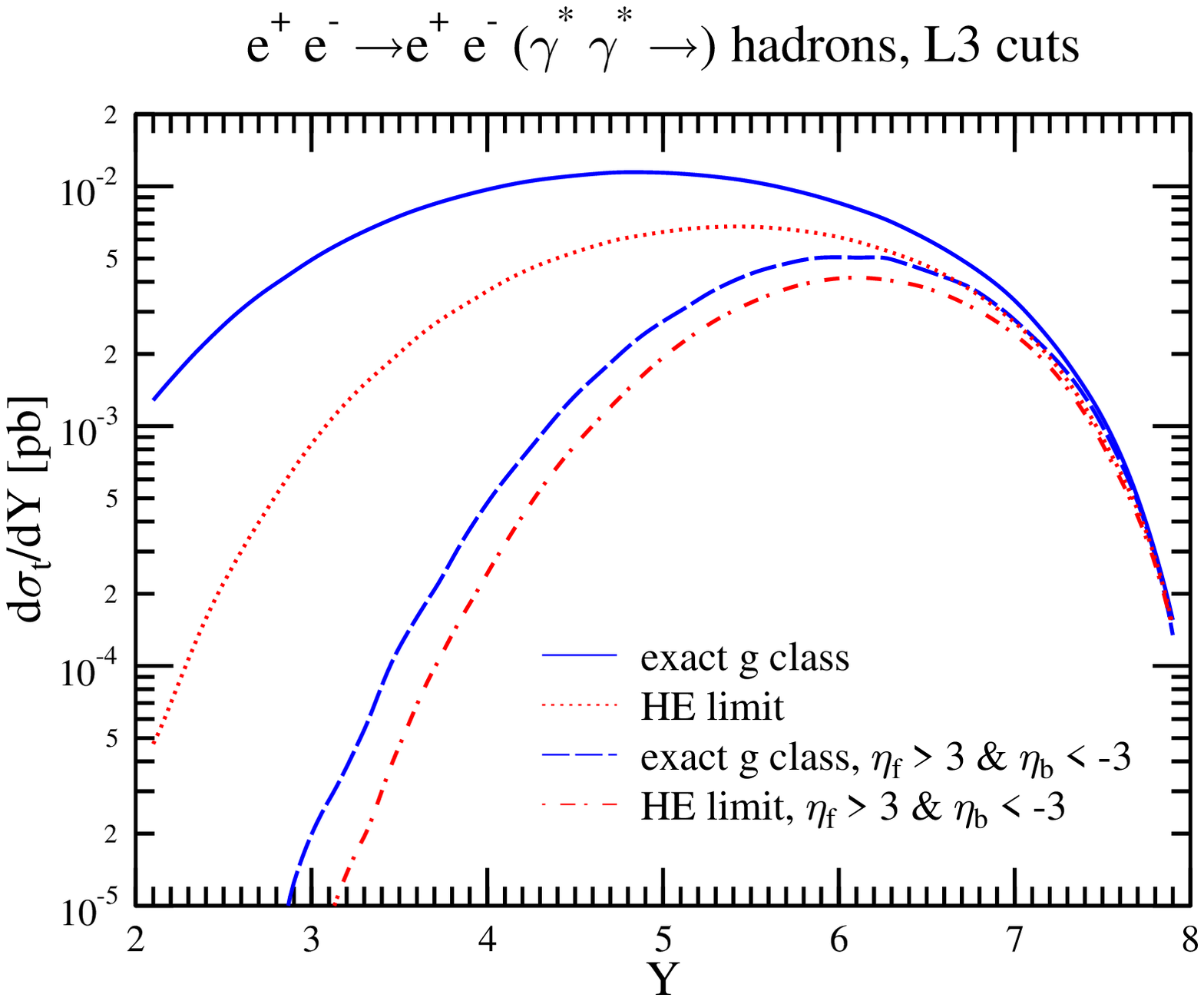,width=14cm}
{The total cross section as a function of $Y$. The solid (dashed) line
represents the contribution of the diagrams of the $g$ class (with a
rapidity cut), the dotted (dot-dashed) line represents the high-energy
limit of the squared matrix element (with a rapidity cut).
\label{fig:HElimit}  }

Figures \ref{fig:yshape} and \ref{fig:etashape} show that in the large $Y$
region, the diagrams of the $g$ class yield an
important contribution to the total cross section. Then it is natural
to ask if, within the acceptance cuts of the LEP2 Collaborations,
the hadronic energy $W$ is sufficiently high to warrant the use of the
high-energy limit (\ref{hekinem}).
We can answer that by comparing the exact contribution
of the diagrams of the $g$ class (\ref{LOcross_distinct})
to the high-energy limit of the squared matrix element
integrated over the exact phase space (\ref{eq:HElimit}).
That comparison is shown in \vfig{fig:HElimit}, where the solid line
is the contribution of the diagrams of the $g$ class and the dotted line
is the high-energy limit of the squared matrix element.
In this and in the following plots, the high-energy limit is obtained in
the $\gamma^*\gamma^*$ centre-of-mass frame.
In order to match the experimental accuracy, in the theoretical prediction
we consider the high-energy limit as accurate only if
the difference between the exact calculation and the high-energy limit
is less than 20\,\%, which is a conservative upper limit of the total
experimental error in the tail of the distributions (see the last bins in
Table \ref{tab:sigee}).
\EPSFIGURE[t]{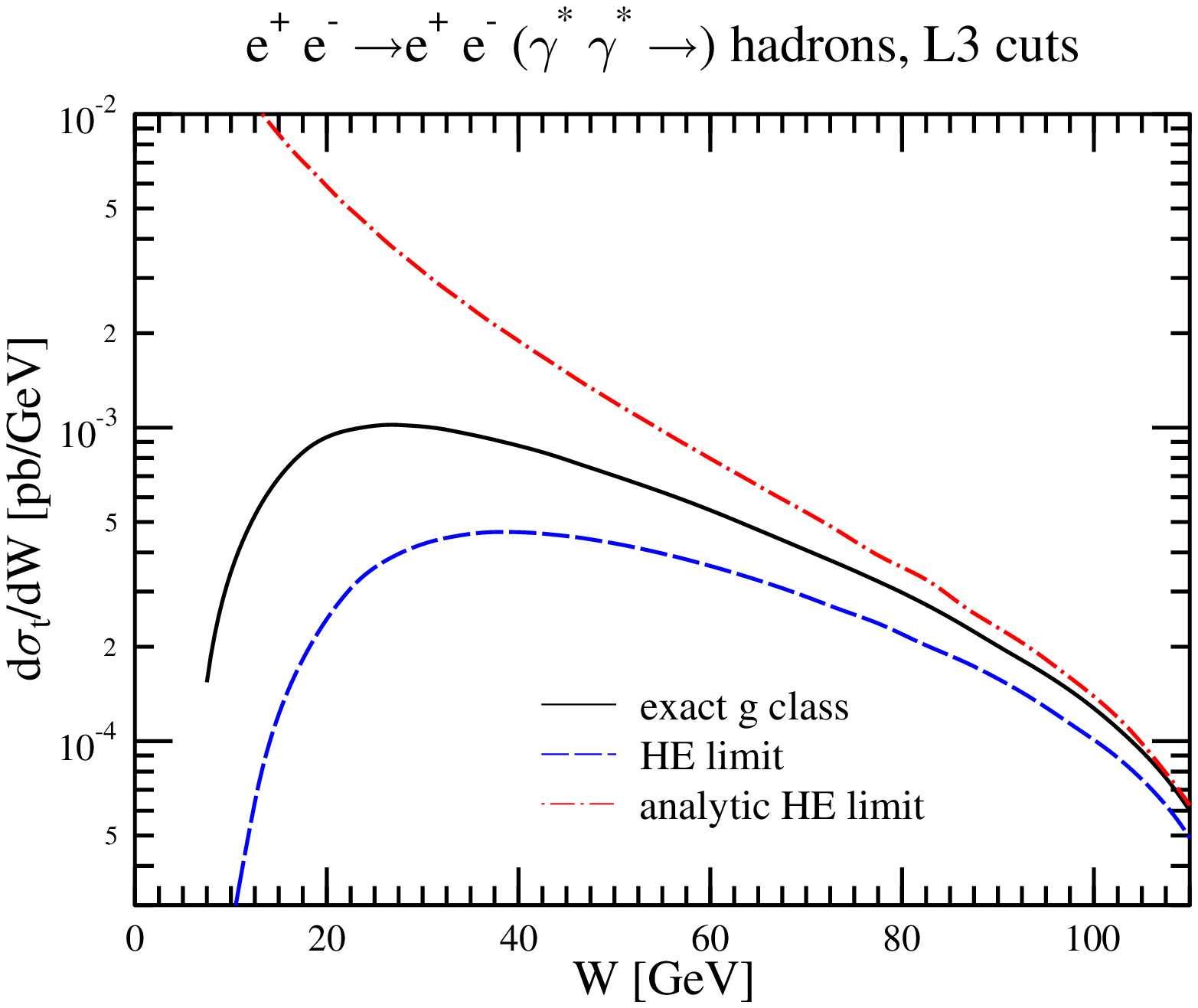,width=14cm}
{The total cross section as a function of the total hadronic energy $W$.
The solid line is the contribution of the diagrams of the $g$ class,
the dashed line and the dot-dashed line are the high-energy limit of
the squared matrix element integrated over the exact phase space
and the high-energy phase space, respectively.
\label{fig:analytic}  }
In \fig{fig:HElimit}
we see that their difference is less than 20\,\% only for $Y\gtrsim 7$.
Unfortunately, the region where $Y\gtrsim 7$ is negligible at the
LEP2 experiments, for the kinematic limit of $Y\approx 8$ is almost
reached, thus the statistics are very small. We can collect the events
in the high-energy region by separating the forward and backward
rapidity regions, which can be achieved by requiring that the sum of the
rapidities of the two most forward momenta, $\eta_f$, be larger than 3 and
that of the other two (backward) momenta, $\eta_b$, be less than $-3$. The
corresponding cross sections are also shown in
\fig{fig:HElimit} for the diagrams of the $g$ class (dashed line) and for
the high-energy limit of the squared matrix element (dot-dashed line).
These curves almost coincide with the cross sections without the
rapidity separation for $Y\gtrsim 7$, but are distinctively smaller for
$Y\lesssim 7$. This is consistent with the difference between the solid
and the dotted lines above. In addition, it shows that for a realistic set-up,
\emph{i.e.} for the cuts of the L3 Collaboration, the high-energy
limit (\ref{hekinem}) is more stringent than the limit
$W^2\gg \mu^2_{\sss W}$, the difference between the two being numerically
negligible only for $Y\gtrsim 7$. Since the high-energy limit (\ref{hekinem})
is the kinematic framework of the BFKL resummation, we conclude that
if a BFKL resummation is used in the $Y\lesssim 7$ region, we expect
the subleading logarithmic corrections to be sizeable.

\EPSFIGURE[t]{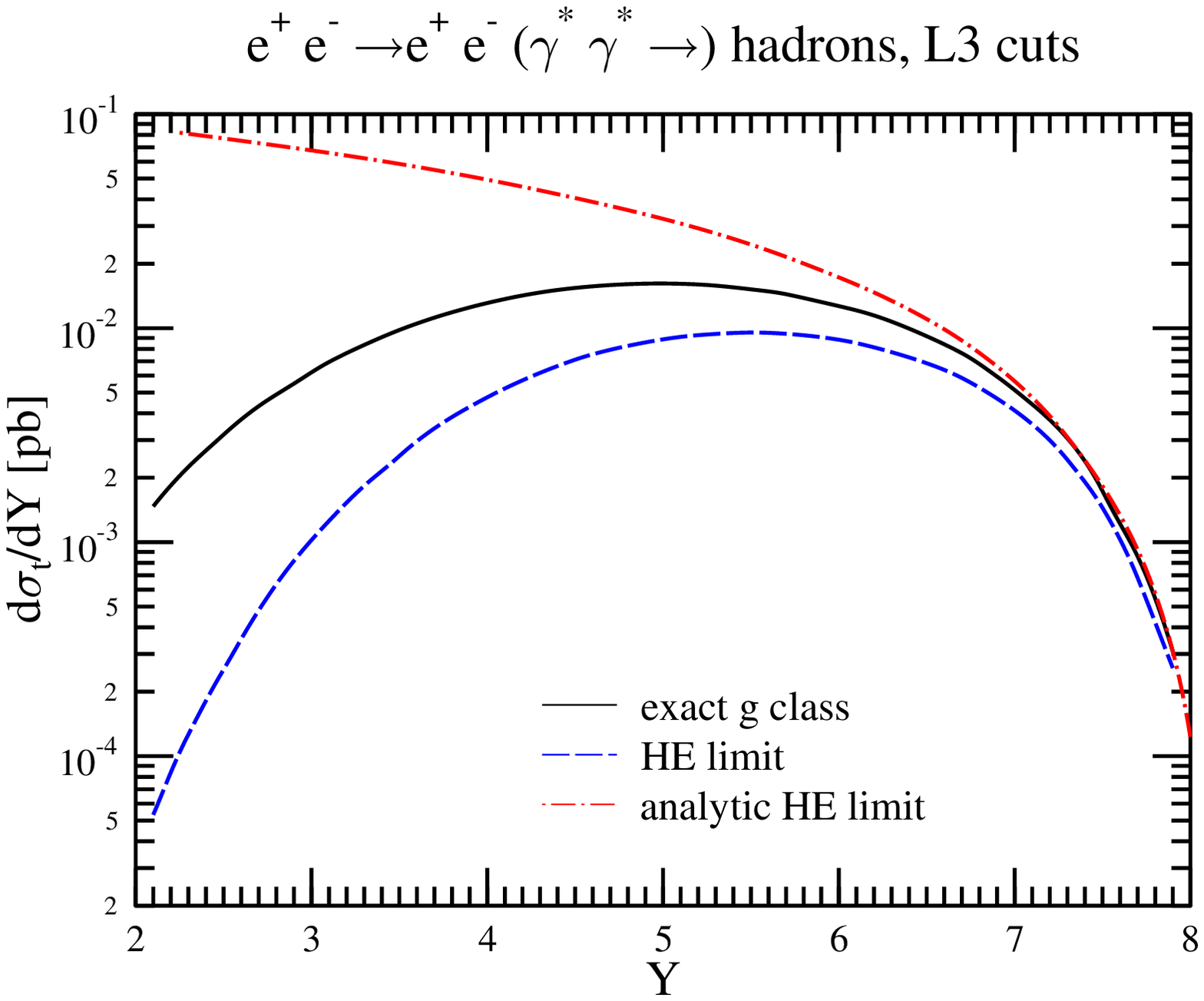,width=14cm}
{Same as Fig.~\ref{fig:analytic}, but with
the total cross section as a function of $Y$.
\label{fig:yanalytic}  }

\EPSFIGURE[t]{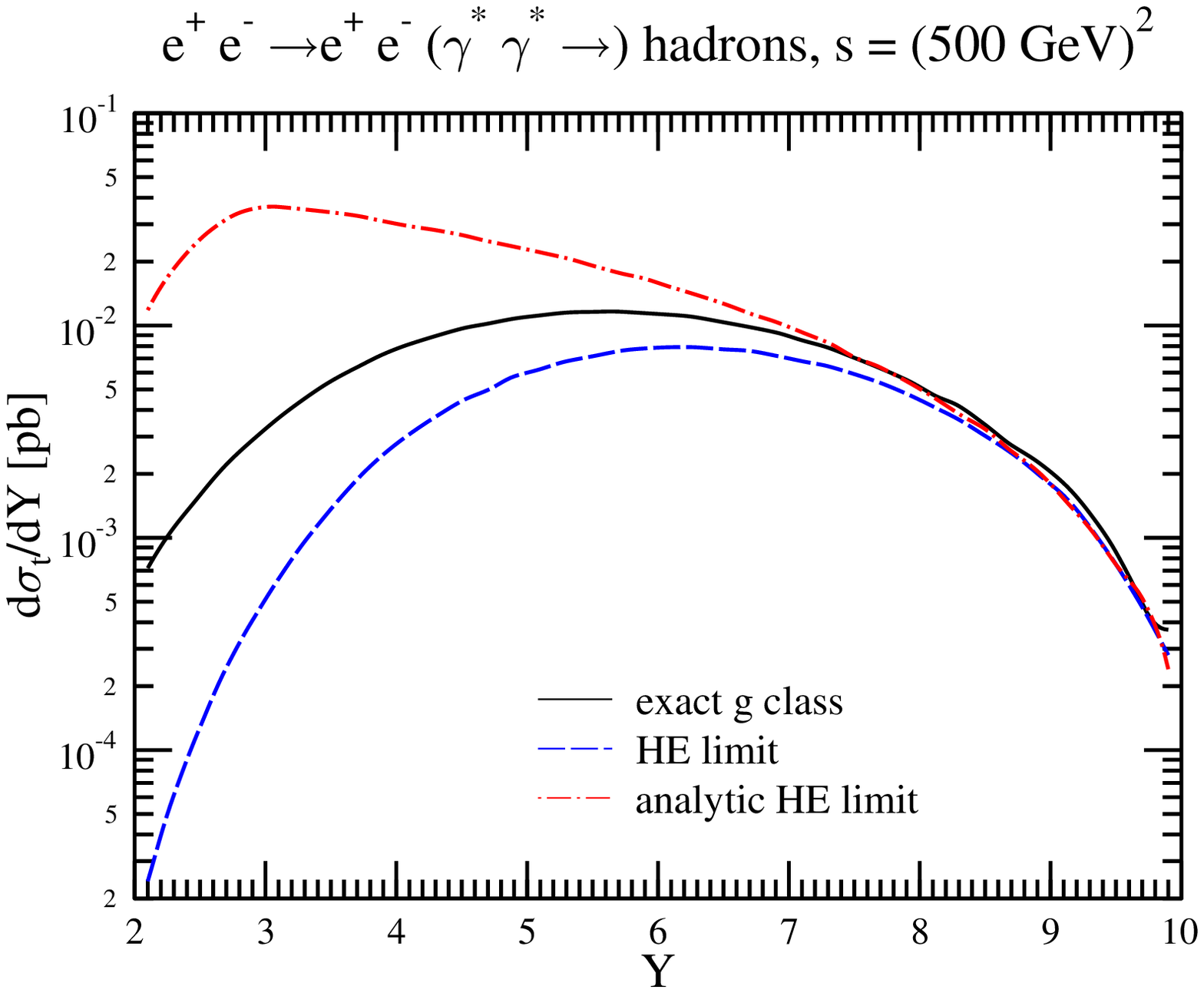,width=14cm}
{Same as Fig.~\ref{fig:yanalytic}, but for a future linear collider
running at $\sqrt{s} = 500$\,GeV.
\label{fig:ynlcanalytic}  }

In \vfig{fig:analytic}, we plot the cross section as a function of
the hadronic energy $W$ using different approximations: the solid line
is the contribution of the diagrams of the $g$ class (\ref{LOcross_distinct}),
the dashed line and the dot-dashed line are the high-energy limit of
the squared matrix element integrated over the exact phase space
(\ref{eq:HElimit}) and the high-energy phase space (\ref{eq:HEPS}),
respectively. In the last one, the limit $s_{\gamma^*\gamma^*}\to\infty$
is taken and the transverse momenta of the quarks are
integrated out analytically. Thus
we shall term it the \emph{analytic} high-energy limit. However, in
this case we cannot use a renormalisation scale like
\eqn{defscale}, which is defined on an event-by-event basis.
Thus, for the sake of comparison, each cross section in \fig{fig:analytic}
was obtained using the scale
\beq
\mu_0^2 = \frac{Q_1^2+Q_2^2}{2}\, .\label{mu0}
\eeq
The three curves converge only for $W \gtrsim 100$\,GeV, which is again
negligible at LEP2 for the poor statistics of the data. In addition, we see
that for $W \lesssim 100$\,GeV the analytic high-energy limit\footnote{In
evaluating the analytic high-energy limit, we used the equivalent photon
approximation for the lepton current (\ref{lepcurrent}), as in
Ref.~\cite{Brodsky:1997sd}.} (dot-dashed line) significantly overestimates the
exact contribution of the diagrams of the $g$ class. For instance, in the [40,
100]\,GeV range, considered by the CERN L3 Collaboration~\cite{Achard:2001kr}
(see the next section), the analytic high-energy limit prediction is about
60\,\% larger than the exact four-quark prediction. In \vfig{fig:yanalytic}, we
plot the cross section as a function of $Y$ using the same approximations as in
\vfig{fig:analytic}. Also the conclusions are basically the same as for
\fig{fig:analytic}, namely the three curves converge only for $Y \gtrsim 7$,
while for $Y < 6$\, the analytic high-energy limit significantly overestimates
the exact contribution of the diagrams of the $g$ class.
Note that the solid and dashed lines of \fig{fig:yanalytic} are the same
as the solid and dotted lines of \fig{fig:HElimit}, but for using the
renormalisation scale (\ref{mu0}) instead of (\ref{defscale}).
\vfig{fig:ynlcanalytic} has the same content as \vfig{fig:yanalytic}, but it is
for a $e^+e^-$ future linear collider running at $\sqrt{s} = 500$\,GeV. For
the sake of illustration, we have taken the following acceptance cuts: the
lepton energies are larger than 40\,GeV, the lepton tagging angles are between
20 and 70\,mrad and the hadronic energy is larger than 20\,GeV. On the photon
virtualities, the cuts above imply that $Q_{1,2}^2\gtrsim 4\ \mbox{GeV}^2$,
while the average virtualities are $\langle Q_i^2 \rangle \simeq
36\,\mbox{GeV}^2$. About the three curves, the same conclusions as for
\fig{fig:yanalytic} can be drawn. However, the much larger statistics (at the
designed luminosity, ${\cal L} = 3.4\cdot
10^{34}$\,cm$^{-2}$s$^{-1}$~\cite{Richard:2001qm}, we expect about 1700 events
in ten days of continuous running) should make also the $Y \gtrsim 7$ region
available to the analysis.

In conclusion, we have considered three successive approximations
to the total cross section at $\ord(\as^2)$:
\vspace*{-.15cm}
\begin{itemize}
  \item the contribution of the diagrams of the $g$ class only,
  \eqn{LOcross_distinct};
\vspace*{-.2cm}
  \item the high-energy limit (\ref{hekinem}) of the squared matrix element,
integrated over:
\vspace*{-.2cm}
  \begin{itemize}
    \item the exact phase space (\ref{eq:HElimit});
\vspace*{-.2cm}
    \item  the high-energy phase space (\ref{eq:HEPS});
  \end{itemize}
\end{itemize}
and we have seen that, although the contributions of the diagrams
with gluon exchange in the crossed channel are numerically important in the
high-$Y$ or high-$W$ regions, in the kinematic range of the
LEP2 experiments the high-energy limit (\ref{hekinem}) is not
sufficiently accurate.

\section{Phenomenological results}
\label{sec:pheno}

In \sec{sec:rapid}, we have analysed the distributions in rapidity of
the final-state partons, and their contribution to the total cross
section in the large $Y$ or large $W$ regions.
As expected, we have found that the
diagrams of the $f$ class yield a contribution which in shape is very
similar to the one of the NLO calculation. Then we may argue that,
if properly counterweighted by the virtual corrections, which at
this moment are unknown, they would yield a rather minor numerical
contribution, since they are an order in $\as$ higher than the NLO one.
On the contrary, the diagrams of the $g$ class, which are by themselves
finite and gauge invariant, have a very different shape in $Y$, becoming
more and more numerically relevant as $Y$ grows. However, we have seen
that in the kinematic range of the LEP2 experiments the diagrams of the
$g$ class must be evaluated exactly, the high-energy limit (\ref{hekinem})
being not sufficiently accurate. Thus in this section
we shall analyse the total cross section as a function of $Y$
through the NLO calculation and/or the diagrams of the $g$ class.

\EPSFIGURE[t]{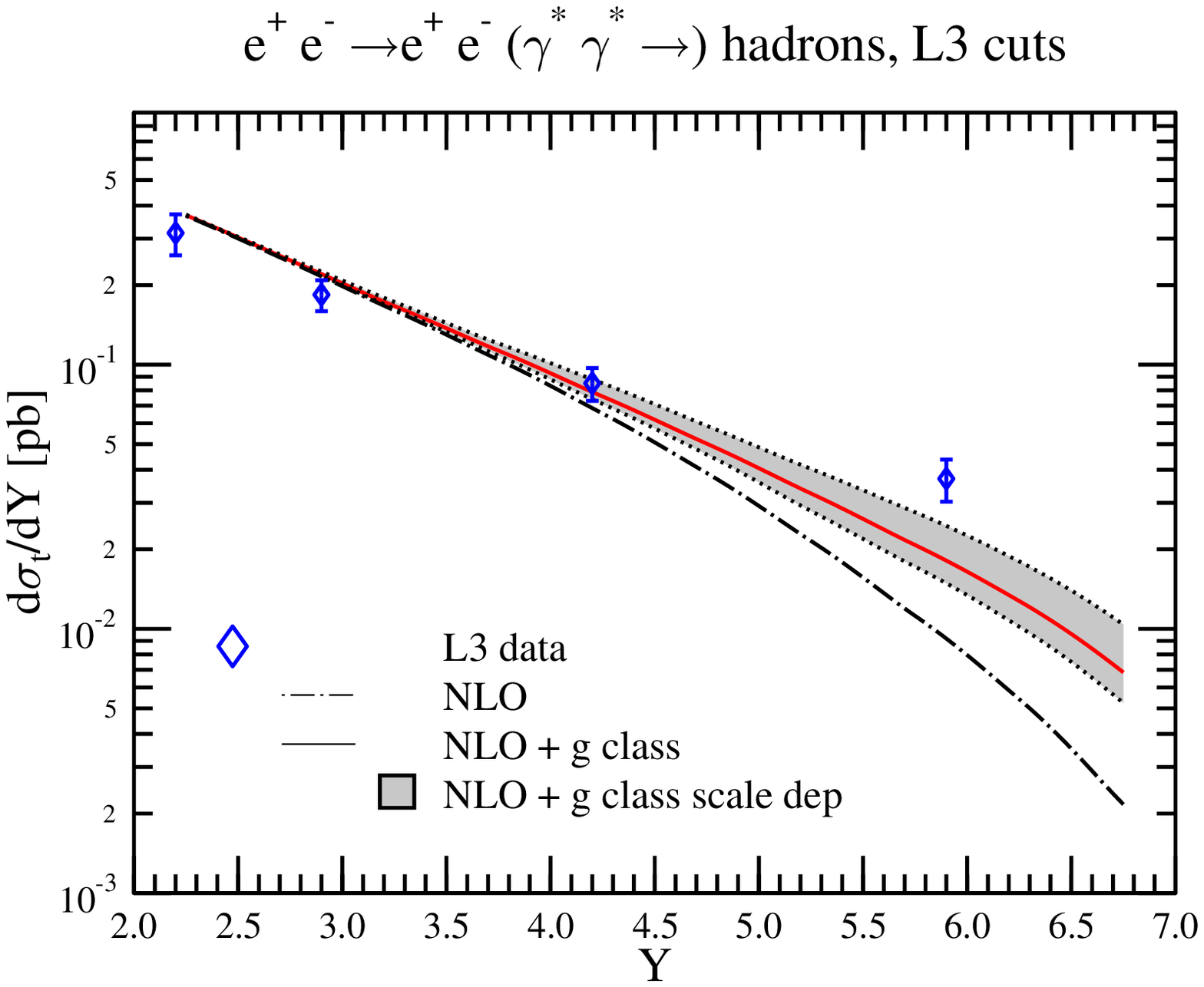,width=14cm} {Total cross section as a function of $Y$,
at NLO (dot-dashed line), and at NLO plus the $\ord(\as^2)$ contribution of the
diagrams of the $g$ class only (solid line). The shaded band has been obtained
by varying the renormalisation scale from $\mu_0/2$ to $2\mu_0$. The points
are the experimental data from the CERN L3 Collaboration~\cite{Achard:2001kr}.
In computing the error bars, we added their statistical and systematic errors
in quadrature. \label{fig:ytotal}  }

In \vfig{fig:ytotal}, we plot the total cross section as a function of $Y$, at
NLO (dot-dashed line) and at NLO plus the $\ord(\as^2)$ contribution of the
diagrams of the $g$ class (solid line). The shaded band has been obtained by
varying the renormalisation scale (\ref{defscale}) $\mu_0/2$ to $2\mu_0$, with
$\mu_0$ given in (\ref{defscale}). The points are the experimental data from
the CERN L3 Collaboration~\cite{Achard:2001kr}. The corresponding numbers are
given in \tab{tab:sigee}. In computing the error bars, we added the statistical
and systematic errors of the data in quadrature. We see that adding the
diagrams of the $g$ class to the NLO calculation decreases the discrepancy
between data and theory at large $Y$, however the data still lie above the
theory prediction, even allowing for a scale uncertainty on the
latter\footnote{We have also computed the total cross section as a function of
$Y$, by using only the diagrams of the $g$ class, {\it i.e.} without the NLO
contribution, and evolving $\as$ with the one-loop running. For the high-$Y$
region ($Y\gtrsim 5$) the outcome is compatible in shape with the dot-dashed
curve, thus it fails as well to describe the data.}. In addition, we remind
that our calculation is performed in the massless limit: no mass effect for
final-state charm and bottom quarks have been included. In
Ref.~\cite{Cacciari:2001cb} it was found the masses to decrease the LO cross
section by 10--15\%. A comparable depletion is expected at NLO. In the case of
four quark production in the analytic high energy limit (defined in the
discussion of Fig.~6), the masses were found to decrease the cross section by
about 20\%~\cite{Bartels:2000sk}. This correction should provide a lower bound
to the exact mass correction in four quark production. Since the four quark
contribution dominates over the NLO calculation at large $Y$, and is negligible
at small $Y$, we should expect the inclusion of the mass corrections to
decrease the solid line of \fig{fig:ytotal} by about 10--15\% at small $Y$ and
by at least 20\% at large $Y$. Therefore the inclusion of the mass dependence
is expected to improve the agreement between data and theory at small $Y$ but
to widen the discrepancy at large $Y$. The same considerations apply to
\vfig{fig:wtotal}, where the total cross section is plotted as a function of
$W$. This was to be expected, since in the large $Y$ limit, $Y$ grows linearly
with the logarithm of $W$.
\EPSFIGURE[t]{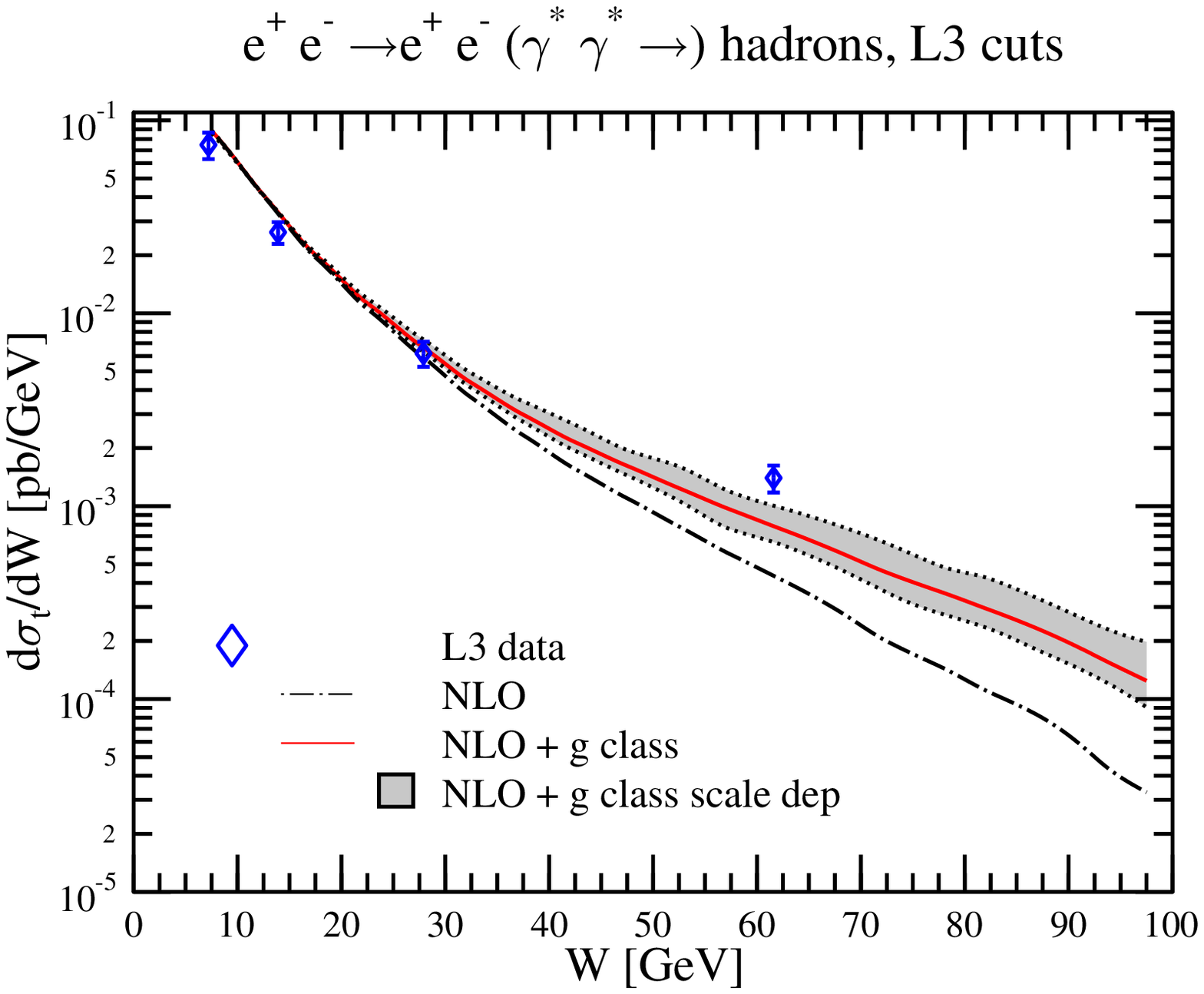,width=14cm}
{Same as in \fig{fig:ytotal}, but with the
total cross section as a function of the hadronic energy $W$.
\label{fig:wtotal}  }

\def\dsigQ{$\mathrm{d\sigma_{\mathrm{ee}} /d}Q^2 $ (pb/GeV$^2$)}
\def\dsigW{$\mathrm{d\sigma_{\mathrm{ee}} /d}W$ (pb/GeV)}
\def\dsigY{$\mathrm{d\sigma_{\mathrm{ee}} /d}Y $ (pb)}
\def\exp#1#2#3{$#1 \pm #2 \pm #3$}
\def\th#1#2#3{$#1 {+#2 \atop -#3}$}

\TABULAR{|c|c|c|c|}
{\hline
$\Delta W$ & L3 data &   NLO   & NLO + $g$ class \\
   (GeV)       & \dsigW  & \dsigW  &     \dsigW      \\
\hline
~5--~10& \exp{0.0747}{0.0096}{0.0067} &
       \th{0.0883}{0.0004}{0.0027} & \th{0.0885}{0.0003}{0.0027} \\
\hline
10--~20& \exp{0.0263}{0.0024}{0.0024} &
       \th{0.0300}{0.0001}{0.0001} & \th{0.0305}{0.0003}{0.0002} \\
\hline
20--~40& \exp{0.0062}{0.0007}{0.0006} &
       \th{0.0057}{0.0001}{0.0003} & \th{0.0064}{0.0006}{0.0003} \\
\hline
40--100& \exp{0.0014}{0.0002}{0.0001} &
       \th{0.0004}{0.0001}{0.0000} & \th{0.0007}{0.0002}{0.0001} \\
\hline \hline
$\Delta Y$ & L3 data &   NLO   & NLO + $g$ class \\
           & \dsigY  & \dsigY  &     \dsigY      \\
\hline
2.0--2.5 & \exp{0.315}{0.048}{0.028} & \th{0.366}{0.001}{0.001} & \th{0.368}{0.002}{0.002} \\
\hline
2.5--3.5 & \exp{0.184}{0.018}{0.017} & \th{0.203}{0.002}{0.001} & \th{0.208}{0.004}{0.003} \\
\hline
3.5--5.0 & \exp{0.085}{0.009}{0.008} & \th{0.070}{0.002}{0.002} & \th{0.080}{0.008}{0.005} \\
\hline
5.0--7.0 & \exp{0.037}{0.006}{0.003} & \th{0.010}{0.001}{0.001} & \th{0.018}{0.006}{0.003} \\
\hline}
{Differential cross sections in $W$ and $Y$ for the process $e^+e^- \to$
hadrons.  For the data the first uncertainty is statistical and the
second systematic.  For the theoretical predictions the error is given
by the renormalization-scale ambiguity.
\label{tab:sigee}}

\section{Conclusions}
\label{sec:conc}

In Ref.~\cite{Cacciari:2001cb}, the question had been addressed of
whether the LEP2 data for the total cross section of
$\gamma^*\gamma^*\to$ hadrons could be described by a NLO calculation.
It was found that the NLO analysis described well the data, except
at the high end of the hadronic energy spectrum.
Through the analysis of the inclusive jet and dijet cross sections,
different kinematic regions were explored, and it was argued that the
region of large $Y$ is particularly susceptible to large logarithms of
type $\ln(W^2/\mu^2_{\sss W})$.

In this work, we have included in the analysis the four parton final states,
which are part of the $\ord(\as^2)$ contribution to the total cross section.
The four--parton final states, which have been included in the massless limit,
contain the diagrams with gluon exchange in the
crossed channel, \emph{i.e.} the diagrams of the $g$ class,
which constitute the leading order of the BFKL resummation.
In \sec{sec:rapid}, we have shown that indeed they play an important
role in the large $Y$ region, however they must be evaluated \emph{exactly}.
In fact, the high-energy limit (\ref{hekinem}), which constitutes the
kinematic framework of the BFKL resummation, is not sufficiently accurate at
LEP2 energies, when compared to the experimental accuracy.
Thus, if a BFKL resummation is used in the large $Y$ region, we expect
the subleading logarithmic corrections to be sizeable.

In \sec{sec:pheno}, we have shown that the contribution of
the diagrams of the $g$ class to the total cross section
reduces the discrepancy between the theory and the LEP2 data of the L3
Collaboration.
However, even allowing for the large scale uncertainty, which is intrinsic
to the diagrams of the $g$ class since they appear for the first time at
$\ord(\as^2)$, the LEP2 data still lie above the theory. We
remind the reader that in the NLO calculation and in the exact four--quark
contribution quark mass effects have not been included.
The inclusion of the mass dependence is expected to
improve the agreement between data and theory at small $Y$ but to
widen the discrepancy at large $Y$.
Thus, in order to describe accurately the data
for $\gamma^*\gamma^*\to$ hadrons, mass effects should be included,
and eventually in the large $Y$ region corrections of
an order higher than $\ord(\as^2)$ should be considered.

\section*{Acknowledgments}

We should like to thank S. Catani, S. Frixione, F. Hautmann, W. Kilgore
and M.L. Mangano for useful discussions.
FM and ZT thank the INFN, sez. di Torino, for its kind
hospitality during the late stage of this work.
This work was supported in part by the EU Fourth Framework Programme
``Training and Mobility of Researchers'', Network ``QCD and particle
structure'', contract FMRX-CT98-0194 (DG 12 - MIHT)
and by the Hungarian Scientific Research Fund grants OTKA T-025482 and
T-038240.

\section*{Note added in proof}

After the completion of this work, we learned that the ALEPH collaboration has
also finished its analysis of double tagged events at LEP to measure the
hadronic cross section in virtual photon-photon scattering~\cite{ALEPH:2002}.
In contradiction to the L3 results, this analysis indicates that the NLO QCD
prediction is sufficient to describe the data in the high-$Y$ region, but is
unable to predict correctly the measured yields in the low-$Y$ region. One has
to resolve this apparent contradiction before applying the analysis presented
in this paper to the ALEPH selection cuts.

\appendix

\section{Chiral-spinor algebra}
\label{sec:spipros}

In order to evaluate the production rates, we use helicity amplitudes,
defined in terms of massless Dirac spinors $\psi_{\pm}(p)$ of fixed helicity,
\begin{equation}
\psi_{\pm}(p) = {1\pm \gamma_5\over 2} \psi(p) \equiv
|p^\pm\rangle\, , \qquad \overline{\psi_{\pm}(p)} \equiv \langle p^\pm|
\, ,\label{spi}
\end{equation}
spinor products,
\begin{equation}
\langle p k\rangle \equiv \langle p^- | k^+ \rangle\, , \qquad
\left[pk\right] \equiv \langle p^+ | k^- \rangle\, ,
\end{equation}
currents,
\bea
\langle i| k | j\rangle &\equiv&
\langle i^-| \slash  \!\!\! k  |j^-\rangle =
\langle i k \rangle \left[k j\right]\, ,\nonumber\\
\langle i| (k+l) | j\rangle &\equiv&
\langle i^-| (\slash  \!\!\! k + \slash  \!\!\! l ) |j^-\rangle
\eea
and Mandelstam invariants
\begin{equation}
s_{pk} = 2p\cdot k =
\langle p k \rangle \left[kp\right]\, , \qquad t_{pkq} =
(p + k + q)^2\, , \qquad t_{pkql} = (p + k + q + l)^2\, .
\end{equation}

\section{Four-quark production}
\label{sec:helamp}

The helicity amplitude featuring two quark pairs and two lepton pairs
is
\beq
\cA_8(1_q,2_\qb;3_\ell,4_\ellb,5_{\ellb'},6_{\ell'};
7_Q,8_\Qb)  = 4 e^4 \gs^2\ T^a_{i_1\ib_2} T^a_{i_7\ib_8}\
A_8(1,2;3,4,5,6;7,8) \, ,\label{app:A8}
\eeq
for two quark pairs of different flavour, and
with $\{1,2\}$ and $\{7,8\}$ the quark pairs\footnote{We normalise
the colour matrices in the fundamental representation as
 tr($T^a T^b$) = $\delta^{ab}$.}, and $\{3,4\}$ and $\{5,6\}$
the lepton pairs. In the colour-stripped sub-amplitude $A_8$, the fermion
flavours, momenta and helicities are implicit in the labels.
$A_8$ is divided into the functions $a_8$, $b_8$ and $c_8$,
\bea
\lefteqn{ A_8(1_q,2_\qb;3_\ell,4_\ellb,5_{\ellb'},6_{\ell'};7_Q,8_\Qb)
= Q_{f_q} Q_{f_Q} a_8(1,2;3,4,5,6;7,8) } \nonumber\\
&& \qquad + Q_{f_q}^2 b_8(1,2;3,4,5,6;7,8) +  Q_{f_Q}^2 c_8(1,2;3,4,5,6;7,8)
\, ,\label{app:a8treesym}
\eea
with $Q_{f_{q(Q)}}$ the electric charge of the quark $q(Q)$ of flavour
$f_{q(Q)}$ and
\bea
a_8(1,2;3,4,5,6;7,8) &=& g_a(1,2;3,4,5,6;7,8) +
g_a(1,2;6,5,4,3;7,8) \nonumber\\ &+&
g_b(1,2;3,4,5,6;7,8) + g_c(1,2;3,4,5,6;7,8)
\nonumber\\ &+& (\{1,2\}\lra\{7,8\})  \nonumber\\
b_8(1,2;3,4,5,6;7,8) &=& f(1,2;3,4,5,6;7,8) + f(1,2;6,5,4,3;7,8)\label{abc}\\
c_8(1,2;3,4,5,6;7,8) &=& b_8(7,8;3,4,5,6;1,2)\, ,\nonumber
\eea
where in the quark pair exchange $(\{1,2\}\lra\{7,8\})$ we swap
the momentum and helicity labels of the (anti)quarks.
In \eqn{app:a8treesym} we have factored the flavour dependence in the quark
electric charges, thus the functions $a_8$, $b_8$ and $c_8$ are
independent of the quark flavours.
In addition, because of the explicit sum in \eqn{abc}
over the different orientations of the quark lines, the labels of the
partons 1 and 7 refer only to quarks, and not to antiquarks.
For distinct flavours each of the functions $a_8$, $b_8$ and $c_8$
is gauge invariant. The functions $g$($f$) refer to
diagrams which feature gluon (quark) exchange in the crossed channel,
\fig{fig:gammastarqqqq}.
The functions $g_b$ and $g_c$ are symmetric under the exchange of the
quark pairs and of the lepton pairs,
\beq
g_i(1,2;3,4,5,6;7,8) = g_i(7,8;6,5,4,3;1,2) \qquad i=b,c \label{exch}
\eeq
For the configuration $(1^-,2^+;3^-,4^+,5^+,6^-;7^-,8^+)$,
the functions $g_a$, $g_b$, $g_c$ and $f$ are,
\bea
g_a &=&
i \, { \spa1.3\spb5.8 \sapp6582 \sapp7134 \over s_{34}\,s_{56}\,t_{134}
t_{568} t_{1234}} \label{treea8}\\
g_b &=&
i \, { \spa1.3\spa6.7\spb2.8 \left( \spb5.6\sapp6134 + \spb5.7\sapp7134
\right) \over s_{34}\,s_{56}\,t_{134}\, t_{567}\, t_{1234}} \label{treeta8}\\
g_c &=&
i \, { \spa1.7\spb2.4\spb5.8 \left( \spa2.3\sapp6582 + \spa4.3\sapp6584
\right) \over s_{34}\,s_{56}\,t_{234}\, t_{568}\, t_{1234}} \label{treeba8}\\
f &=& i \, \Biggl(
{ \spa1.3\spb2.5 \sapp6258 \sapp7134 \over s_{34}\,s_{56}\,s_{78}\,
t_{134}\,t_{256} } \nonumber\\ && +\,
{ \spa1.7\spb2.5 \sapp6254 \sapp3178 \over
s_{34}\,s_{56}\,s_{78}\, t_{178}\,t_{256} } \label{treeb8}\\ && +\,
{ \spa1.3\spb2.8 \sapp7285 \sapp6134 \over s_{34}\,s_{56}\,s_{78}\,
t_{134}\,t_{278} } \Biggr) \, .\nonumber
\eea
For all of the other helicity configurations, the functions $g_a$, $g_b$,
$g_c$ and $f$ assume a functional form which is in principle different,
however, the other lepton-helicity configurations are simply obtained
by exchanging the labels 3 and 4 and/or 5 and 6 in \eqn{app:A8}. Analogously,
we show in \app{sec:appa} that the other quark-helicity configurations are
obtained by exchanging the labels 1 and 2 and/or 7 and 8.

In the squared amplitude, the sum over distinct
flavours can be written as
\bea
\lefteqn{ \frac12 \sum_{f_q,f_Q\ne f_q} |A_8(1,2;3,4,5,6;7,8)|^2 }
\nonumber\\ &=&
\frac12\Bigg[
\sum_{f_q,f_Q\ne f_q} \left\{ Q_{f_q}^2 Q_{f_Q}^2 \left[ |a_8|^2 + 2{\rm Re}
(b_8^*c_8) \right] + 2 Q_{f_q}^3 Q_{f_Q} {\rm Re}(a_8^*b_8) +
2 Q_{f_q} Q_{f_Q}^3 {\rm Re}(a_8^*c_8) \right\} \nonumber\\ &+&
(n_f-1) \biggl( \sum_f Q_f^4 \biggr) \left( |b_8|^2 + |c_8|^2 \right)
\Bigg] \, ,\label{sums}
\eea
with $n_f$ the number of quark flavours, and
\bea
&& \sum_f Q_f^4 = Q_u^4 n_u + Q_d^4 n_d \\ && \hspace{-0.4cm}
\sum_{f_q,f_Q\ne f_q} \hspace{-0.2cm} Q_{f_q}^i Q_{f_Q}^j =
Q_u^iQ_u^j n_u(n_u-1) +
Q_d^iQ_d^j n_d(n_d-1) + \left(Q_u^iQ_d^j + Q_u^jQ_d^i \right) n_u n_d
\nonumber
\eea
with $i,j$ any integer power, and with $Q_u=2/3$, $Q_d=-1/3$ and $n_{u(d)}$
the number of up(down)-type quarks. In \eqn{sums} the factor of $1/2$
appears in
order to avoid double counting, since the amplitudes are symmetric with
respect to the interchange of the two quark lines.

For two quark pairs of equal flavour, the sub-amplitude $A_8$ of
\eqn{app:a8treesym} becomes
\beq
A_8^\id(1_q,2_\qb;3_\ell,4_\ellb,5_{\ellb'},6_{\ell'};7_q,8_\qb)
= Q_f^2  a_8^\id(1,2;3,4,5,6;7,8)\, ,\label{a8treeid}
\eeq
with
\bea
a_8^\id(1,2;3,4,5,6;7,8) &=& g_a(1,2;3,4,5,6;7,8) +
g_a(1,2;6,5,4,3;7,8) \nonumber\\ &+&
g_b(1,2;3,4,5,6;7,8) + g_c(1,2;3,4,5,6;7,8) \label{aid}\\
&+& f(1,2;3,4,5,6;7,8) + f(1,2;6,5,4,3;7,8) \nonumber\\ &+&
(\{1,2\}\lra\{7,8\})\, .\nonumber
\eea
Note that in this instance the diagrams with gluon exchange in the
crossed channel, corresponding to the functions $g_a$, $g_b$ and $g_c$,
are in the same gauge class, while those featuring
quark exchange in the crossed channel form a different gauge class.
In addition, quarks of equal flavour are indistinguishable,
thus we must add to \eqn{app:A8} the contribution with the quarks (but not the
anti-quarks) exchanged, and antisymmetrise the whole amplitude
in the colour and momentum labels,
\bea
&&
\!\!\!\!\!\!\!\!
\cA_8^\id(1_q,2_\qb;3_\ell,4_\ellb,5_{\ellb'},6_{\ell'};
7_q,8_\qb) = 4 e^4 \gs^2
\nonumber\\ &&
\!\!\!\!\!\!\!\!
\times {1\over 2} \Biggl\{
 \left[ (T^a)_{i_1\ib_2} (T^a)_{i_7\ib_8} +
(T^a)_{i_7\ib_2} (T^a)_{i_1\ib_8} \right]
\left[ A_8^\id(1,2;3,4,5,6;7,8) - A_8^\id(7,2;3,4,5,6;1,8)\right]
\nonumber\\ &&
\!\!
+ \left[ (T^a)_{i_1\ib_2} (T^a)_{i_7\ib_8} -
(T^a)_{i_7\ib_2} (T^a)_{i_1\ib_8} \right]
\left[ A_8^\id(1,2;3,4,5,6;7,8) + A_8^\id(7,2;3,4,5,6;1,8)\right] \Biggr\}
\nonumber\\ &=& 4 e^4 \gs^2 \left[
(T^a)_{i_1\ib_2} (T^a)_{i_7\ib_8} A_8^\id(1,2;3,4,5,6;7,8) -
(T^a)_{i_7\ib_2} (T^a)_{i_1\ib_8} A_8^\id(7,2;3,4,5,6;1,8)\right]\,
,\nonumber\\
\label{A8id}
\eea
i.e. we can antisymmetrise \eqns{app:A8}{a8treeid} by subtracting the same
expression with the colour and momentum labels of the quarks exchanged.

In crossing to the physical region,
we choose 4 as the incoming electron and 6 as the incoming positron.
For a fixed lepton-helicity configuration,
e.g.~$(3_\ell^-,4_\ellb^+,5_{\ellb'}^+,6_{\ell'}^-)$, the production
rate is obtained by summing over the quark-helicity configurations,
\bea
\lefteqn{ d\sigma(3_\ell^-,4_\ellb^+,5_{\ellb'}^+,6_{\ell'}^-) =
{1\over 2s} d\cP_6  16 \left(N_c^2 -1\right) (4\pi\aem )^4 (4\pi\as)^2 }
\nonumber\\ &\times& \Biggl\{ {1\over 2}
\sum_{f_q,f_Q\ne f_q} \Bigl[ |A_8(1^-,2^+;3^-,4^+,5^+,6^-;7^-,8^+)|^2
+ |A_8(2^-,1^+;3^-,4^+,5^+,6^-;7^-,8^+)|^2 \nonumber\\ && \quad
+\, |A_8(1^-,2^+;3^-,4^+,5^+,6^-;8^-,7^+)|^2 +
|A_8(2^-,1^+;3^-,4^+,5^+,6^-;8^-,7^+)|^2 \Bigr]
\nonumber\\ &+&
\frac14
 \Bigl(\sum_f Q_f^4\Bigr) \Bigl[ |a_8^\id(1^-,2^+;3^-,4^+,5^+,6^-;7^-,8^+)|^2
 + |a_8^\id(7^-,2^+;3^-,4^+,5^+,6^-;1^-,8^+)|^2 \nonumber\\ && \;
 + {2\over N_c} {\rm Re} \left[a_8^\id(1^-,2^+;3^-,4^+,5^+,6^-;7^-,8^+)^*
 a_8^\id(7^-,2^+;3^-,4^+,5^+,6^-;1^-,8^+)\right] \nonumber\\ && \;
 + |a_8^\id(2^-,1^+;3^-,4^+,5^+,6^-;7^-,8^+)|^2
 + |a_8^\id(1^-,2^+;3^-,4^+,5^+,6^-;8^-,7^+)|^2 \nonumber\\ && \; +
 |a_8^\id(7^-,2^+;3^-,4^+,5^+,6^-;8^-,1^+)|^2
 + |a_8^\id(2^-,7^+;3^-,4^+,5^+,6^-;1^-,8^+)|^2 \nonumber\\ && \; +
 |a_8^\id(2^-,1^+;3^-,4^+,5^+,6^-;8^-,7^+)|^2 +
 |a_8^\id(2^-,7^+;3^-,4^+,5^+,6^-;8^-,1^+)|^2 \nonumber\\ && \;
 + {2\over N_c} {\rm Re}\, \bigl[ a_8^\id(2^-,1^+;3^-,4^+,5^+,6^-;8^-,7^+)^*
 a_8^\id(2^-,7^+;3^-,4^+,5^+,6^-;8^-,1^+)\bigr]\Bigr] \Biggr\}\, ,\nonumber\\
\label{LOcross}
\eea
where $d\cP_6$ is the phase space for the $e^+e^- q\bar q Q\bar Q$
final state,
\beq
d{\cal P}_6 = \prod_{i} {d^3 p_i\over (2\pi)^3 2E_i}\,
(2\pi)^4 \, \delta^4(p_4 + p_6 - p_1 - p_2 - p_3 - p_5 - p_7 - p_8)\,
,\label{6phase}
\eeq
with $i$ = 1, 2, 3, 5, 7, 8.
In \eqn{LOcross} we have performed explicitly
the sum over quarks of equal flavour, and we have multiplied by the
symmetry factor 1/4 for two identical quarks and two identical antiquarks.
The sum over quarks of different flavour is performed in \eqn{sums}.
The unpolarised rate is given by averaging over the rates for the four
lepton-helicity configurations.

\subsection{Symmetries under helicity flips of the quark lines}
\label{sec:appa}

Symmetry relations between the functions $g_a$, $g_b$, $g_c$ and $f$
with respect to the configuration $(1^-,2^+;3^-,4^+,5^+,6^-;7^-,8^+)$,

$(a)$ under helicity flip of the pair $\{1,2\}$,
\bea
g_a(1^+,2^-;3^-,4^+,5^+,6^-;7^-,8^+) &=& - g_c(2^-,1^+;3^-,4^+,5^+,6^-;7^-,8^+)
\nonumber\\
g_a(7^-,8^+;3^-,4^+,5^+,6^-;1^+,2^-) &=& - g_b(2^-,1^+;6^-,5^+,4^+,3^-;7^-,8^+)
\nonumber\\
g_b(1^+,2^-;3^-,4^+,5^+,6^-;7^-,8^+) &=& - g_a(7^-,8^+;6^-,5^+,4^+,3^-;2^-,1^+)
\label{1+2-}\\
g_c(1^+,2^-;3^-,4^+,5^+,6^-;7^-,8^+) &=& - g_a(2^-,1^+;3^-,4^+,5^+,6^-;7^-,8^+)
\nonumber\\
f(1^+,2^-;3^-,4^+,5^+,6^-;7^-,8^+) &=& - f(2^-,1^+;6^-,5^+,4^+,3^-;7^-,8^+)
\nonumber\\
f(7^-,8^+;3^-,4^+,5^+,6^-;1^+,2^-) &=& - f(7^-,8^+;3^-,4^+,5^+,6^-;2^-,1^+)
\nonumber
\eea

$(b)$ under helicity flip of the pair $\{7,8\}$,
\bea
g_a(1^-,2^+;3^-,4^+,5^+,6^-;7^+,8^-) &=& - g_b(1^-,2^+;3^-,4^+,5^+,6^-;8^-,7^+)
\nonumber\\
g_a(7^+,8^-;3^-,4^+,5^+,6^-;1^-,2^+) &=& - g_c(1^-,2^+;6^-,5^+,4^+,3^-;8^-,7^+)
\nonumber\\
g_b(1^-,2^+;3^-,4^+,5^+,6^-;7^+,8^-) &=& - g_a(1^-,2^+;3^-,4^+,5^+,6^-;8^-,7^+)
\label{7+8-}\\
g_c(1^-,2^+;3^-,4^+,5^+,6^-;7^+,8^-) &=& - g_a(8^-,7^+;6^-,5^+,4^+,3^-;1^-,2^+)
\nonumber\\
f(1^-,2^+;3^-,4^+,5^+,6^-;7^+,8^-) &=& - f(1^-,2^+;3^-,4^+,5^+,6^-;8^-,7^+)
\nonumber\\
f(7^+,8^-;3^-,4^+,5^+,6^-;1^-,2^+) &=& - f(8^-,7^+;6^-,5^+,4^+,3^-;1^-,2^+)
\nonumber
\eea

$(c)$ under helicity flips of the pairs $\{1,2\}$ and $\{7,8\}$,
reflection
\bea
g_a(1^+,2^-;3^-,4^+,5^+,6^-;7^+,8^-) &=& g_a(8^-,7^+;6^-,5^+,4^+,3^-;2^-,1^+)
\nonumber\\
g_a(7^+,8^-;3^-,4^+,5^+,6^-;1^+,2^-) &=& g_a(2^-,1^+;6^-,5^+,4^+,3^-;8^-,7^+)
\nonumber\\
g_b(1^+,2^-;3^-,4^+,5^+,6^-;7^+,8^-) &=& g_c(2^-,1^+;3^-,4^+,5^+,6^-;8^-,7^+)
\label{1+2-7+8-}\\
g_c(1^+,2^-;3^-,4^+,5^+,6^-;7^+,8^-) &=& g_b(2^-,1^+;3^-,4^+,5^+,6^-;8^-,7^+)
\nonumber\\
f(1^+,2^-;3^-,4^+,5^+,6^-;7^+,8^-) &=& f(2^-,1^+;6^-,5^+,4^+,3^-;8^-,7^+)
\nonumber\\
f(7^+,8^-;3^-,4^+,5^+,6^-;1^+,2^-) &=& f(8^-,7^+;6^-,5^+,4^+,3^-;2^-,1^+)
\nonumber
\eea

\eqnss{1+2-}{1+2-7+8-} show that the helicity flip of the
pairs $\{1,2\}$ and/or $\{7,8\}$ reshuffles the functions in \eqn{a8treesym}
however it does not change their sum but only the overall sign, thus the
helicity flip of the
pairs $\{1,2\}$ and/or $\{7,8\}$ of the sub-amplitude (\ref{a8treesym}) can
be achieved by merely exchanging the corresponding labels.

\section{The impact factor for $\boldsymbol{\gamma^* g^* \to q \bar{q} }$}
\label{sec:dif}

In order to derive the impact factor for $\gamma^* g^* \to q \bar{q}$, the
simplest is to use either the amplitudes for $e^+e^-\to q \bar{q} g g$
or those for $e^+e^-\to q \bar{q} Q \bar{Q}$. They are collected, for
instance, in Ref.~\cite{Bern:1998sc}. We shall take a lepton and a
parton (quark or gluon), of momenta $p_a$ and $p_b$ respectively as
the incoming particles, and a lepton, a quark pair and a parton
of momenta $p_{a'}$, $p_1$, $p_2$ and $p_{b'}$ as the outgoing
particles. All momenta are taken as outgoing,
so that momentum conservation reads $p_a+p_b+p_{a'}+p_1+p_2+p_{b'}=0$.
We use light-cone coordinates: $p^\pm = p^0 \pm p^z$ and
complex transverse coordinates $p_{\perp} = p^x + i p^y$, and write a
four vector $p^\mu=(p^0,p^1,p^2,p^3)$ as
$(p^+,p^-;p_{\perp}, \bar p_{\perp})$, where
$\bar p_{\perp} = p^x - i p^y$. With this notation the scalar
product is $2 p\cdot q =
p^+q^- + p^-q^+ - p_{\perp} \bar q_{\perp} - \bar p_{\perp} q_{\perp}$.
For real four vectors, \emph{e.g.} for momenta, for which
$\bar p_{\perp} = p_{\perp}^*$, we shall use the shorter
notation $p = (p^+,p^-;p_{\perp})$.

We define the photon momentum to be $k = - p_a - p_{a'}$, and choose as
the reference frame the virtual photon-parton frame, defined by
$p_b = (0, p_b^-; 0_\perp)$ and $k = (k^+, k^-; 0_\perp)$.
Momentum conservation requires that
\bea
k^+ &=& p_1^+ + p_2^+ + p_{b'}^+ \nonumber\\
k^- &=& p_1^- + p_2^- + p_b^- + p_{b'}^- \label{kin}\\
0 &=& p_{1\perp} + p_{2\perp} + p_{b'\perp} \nonumber
\eea
Now we take the high-energy limit, where the outgoing partons
are strongly ordered on the light cone and have comparable transverse momentum,
\begin{equation}
p_1^+ \simeq p_2^+ \gg p_{b'}^+;\qquad p_1^- \simeq p_2^- \ll p_{b'}^-;\qquad
|p_{1\perp}| \simeq |p_{2\perp}| \simeq |p_{b'\perp}|\, .\nonumber
\end{equation}
While the transverse components of \eqn{kin} remain untouched, the
light-cone components are approximated by $k^+\simeq p_1^+ + p_2^+$ and
$k^-\simeq p_b^- + p_{b'}^-$.
In addition, the virtual photon-parton centre-of-mass energy
\beq
s_{\gamma^* p} = (k-p_b)^2 = k^+ k^- - k^+ p_b^- \simeq k^+ p_{b'}^-
\eeq
is required in the high-energy limit to be much larger than the virtual photon
momentum transfer $k^2$, i.e. $s_{\gamma^* p} \gg |k^2|$.
This entails that $p_{b'}^- \gg |k^-|$. Thus in the momentum conservation
along the minus direction
the momentum $k^-$ can be neglected, and we can summarise
the momentum conservation in the high-energy limit as
\bea
- p_a^+ - p_{a'}^+ = k^+ &\simeq& p_1^+ + p_2^+ \nonumber\\
- p_b^- &\simeq& p_{b'}^- \label{mrkin}\\
0 &=& p_{1\perp} + p_{2\perp} + p_{b'\perp}\, .\nonumber
\eea
\eqn{mrkin} can be viewed as defining
two scattering centres through the $+$ and $-$ momentum conservation,
which act independently. The two scattering centres are
linked by the transverse momentum conservation only.

Next, we must approximate the exact amplitudes in the high-energy limit.
The colour decomposition of the amplitude for $e^+e^-\to q \bar{q} g g$
in the conventions of Ref.~\cite{Bern:1998sc}, is
\beq
\cA_6(1_q,b,b',2_\qb;a_\eb,{a'}_e) = -2 e^2 \gs^2 Q_{f_q} \sum_{\sigma\in S_2}
(T^{a_{\sigma(b)}} T^{a_{\sigma(b')}} )_{i_1\ib_2}
A_6(1_q,\sigma(b),\sigma(b'),2_\qb;a_\eb,a'_e) \label{color}
\eeq
with sub-amplitudes\footnote{We use the same sub-amplitudes as in
Ref.~\cite{Bern:1998sc}, but we ignore the overall factor of $i$. In addition,
we have neglected the configuration with like-helicity gluons, since it
subleading in the high-energy limit.}
\bea
&&
A_6(1^+,b^+,{b'}^-,2^-,a^-,{a'}^+) = {1\over s_{b{b'}} s_{a{a'}} }
\biggl[ {\spa{b'}.1 \spb1.b \spa2.a \langle b'|(1+b)|a'\rangle
\over \spa1.b t_{1bb'} } \label{bdk1}\\ &&\qquad\qquad -
{\spa{b'}.2 \spb2.b \spb1.{a'} \langle a|(b'+2)| b\rangle
\over \spb{b'}.2 t_{b{b'}2} }
- { \langle a|(b'+2)| b\rangle  \langle b'|(1+b)|a'\rangle
\over \spa1.b \spb{b'}.2} \Bigr]\, ,\nonumber
\eea
\bea
A_6(1^+,b^-,{b'}^+,2^-,a^-,{a'}^+) &=& {1\over s_{b{b'}} s_{a{a'}} }
\Bigl[ - {\spb1.{b'}^2 \spa2.a
\langle b|(1+b')|a'\rangle \over \spb1.b t_{1b{b'}} }
\label{bdk2}\\ &+& {\spa{b}.2^2 \spb1.{a'}
\langle a|(b+2)|b'\rangle \over \spa{b'}.2 t_{bb'2} }
+ { \spb1.{b'} \spa{b}.2 \spb1.{a'} \spa2.a \over \spb1.b \spa{b'}.2 }
\Bigr]\, .\nonumber
\eea
The sub-amplitudes (\ref{bdk1}) and (\ref{bdk2}) are symmetric under
the exchange
\begin{equation}
1\lra 2\, ,\quad b\lra {b'}\, ,\quad a\lra {a'}\, ,\quad \langle ij\rangle\lra
[ji]\, .\label{flip}
\end{equation}
Alternatively, the amplitude for $e^+e^-\to q \bar{q} Q \bar{Q}$ can
be used,
\bea
\label{bdk6}
\lefteqn{ {\cal A}_6^{\rm tree}(1_q,b_\Qb,b'_Q,2_\qb;a_{\bar e},{a'}_e)=
-2\,e^2\,\gs^2\, \sum_{a=1}^8 T^a_{i_1\bar\imath_2}T^a_{i_{b'}\bar\imath_b} }
\\ \nn &\times&
\left[Q_{f_q}\,A_6^{\rm tree}(1_q,b_\Qb,{b'}_Q,2_\qb;a_{\bar e},{a'}_e)
     +Q_{f_Q}\,A_6^{\rm tree}({b'}_Q,2_\qb,1_q,b_\Qb;a_{\bar e},{a'}_e)\right]
\eea
with sub-amplitude
\bea
\label{A6PPmmmp}
\lefteqn{ A_6^{\rm tree}(1^+,b^+,{b'}^-,2^-;a^-,{a'}^+) } \nn\\
&=& {1\over s_{b{b'}}s_{a{a'}} } \left[
\frac{\spb1.b \spa{a}.2 \sapp{b'}1b{a'}}{t_{1b{b'}}}
+\frac{\spa{b'}.2 \spb{a'}.1 \sapp{a}{b'}2b}{t_{b{b'}2}} \right] \:.
\eea
In order to evaluate the sub-amplitudes above, we need to compute the
Mandelstam invariants and the spinor products in the high-energy limit.
In the virtual photon-parton frame, the three-particle invariants can be
written {\it exactly} as
\beq
t_{ia{a'}} = (p_i+p_a+p_{a'})^2 =
k^2 - {k^+\over p_i^+} |p_{i_\perp}|^2 -
{p_i^+\over k^+} k^2\, ,\qquad i=1,2\, ,\label{tinvpp}
\eeq
where $x_i = p_i^+/k^+$, with $i=1,2$, are the momentum fractions
of the final-state quarks with respect to the virtual photon.
In the high-energy limit, we can fix $x_1 = x$ and $x_2 = 1 -x$, and
rewrite the invariants (\ref{tinvpp}) as
\bea
t_{1aa'} &=& t_{2bb'} = (1-x) k^2 - {|p_{1\perp}|^2\over x}\, ,\nonumber\\
t_{2aa'} &=& t_{1bb'} = x k^2 - {|p_{2\perp}|^2\over (1-x)}\,
.\label{tinvhe}
\eea
In light-cone coordinates, a generic spinor product can be written as
\beq
\langle p_i p_j\rangle = p_{i\perp}\sqrt{p_j^+\over p_i^+} - p_{j\perp}
\sqrt{p_i^+\over p_j^+}\, .
\eeq
In the kinematics (\ref{mrkin}), the spinor products are
\begin{eqnarray}
\langle p_a p_b\rangle &=&
-\sqrt{\hat s} \simeq - \sqrt{(p_1^+ + p_2^+ + p_{a'}^+) p_{b'}^-}\,
,\nonumber\\
\langle p_a p_{b'}\rangle &=&
-i \sqrt{-p_a^+\over p_{b'}^+}\, p_{b'_\perp} \simeq i
{p_{b'_\perp}\over |p_{b'_\perp}|} \langle p_a p_b\rangle\, ,\nn\\
\langle p_a p_j\rangle &=& i \left( (-p_{a_\perp}) \sqrt{p_j^+\over -p_a^+}
- p_{j_\perp} \sqrt{-p_a^+\over p_j^+}\right)\, ,\label{spipro}\\
\langle p_j p_b\rangle &=& i \sqrt{-p_b^- p_j^+}\,
\simeq i \sqrt{p_j^+ p_{b'}^-}\, ,\nonumber\\
\langle p_{b'} p_b\rangle &=& i \sqrt{-p_b^- p_{b'}^+}\,
\simeq i |p_{b'_\perp}|\, ,\nonumber\\
\langle p_j p_{b'}\rangle &=&
p_{j_\perp}\sqrt{p_{b'}^+\over p_j^+} - p_{b'_\perp}
\sqrt{p_j^+\over p_{b'}^+} \simeq - p_{b'_\perp}\,
\sqrt{p_j^+\over p_{b'}^+}\, ,\nonumber
\end{eqnarray}
with $j=1,2,a'$, and where we have
taken the phase conventions of Ref.~\cite{DelDuca:2000ha}.
Momentum conservation (\ref{mrkin}) implies that $p_a^+ = - p_{a'}^+ -
p_1^+ - p_2^+$ and $p_{a\perp} = - p_{a'\perp}$.

Using the invariants (\ref{tinvhe}) and the spinor products (\ref{spipro}),
the high-energy expansion of \eqn{color} can be written as
\bea
\lefteqn{\cA_6(1_q^\nu,b^\rho,{b'}^{\rho'},2_\qb^{-\nu},a_\eb^{-\lambda},
{a'}_e^\lambda) }  \nonumber \\ &&
= 2\, s_{\gamma^* p} \left[ \gs e^2 Q_{f_q}\, T^c_{i_1\ib_2}\,
\sqrt{2} V_f(a_\eb^{-\lambda};{a'}_e^\lambda,1_q^\nu,2_\qb^{-\nu}) \right]
{1\over t} \left[\frac{i}{\sqrt{2}} \gs f^{b{b'}c}\,
C(b_g^{\rho};{b'}_g^{\rho'}) \right]\, ,\label{hecolpp}
\eea
where $s_{\gamma^* p}\simeq k^+ p_{b'}^-$ is the virtual photon-parton
centre-of-mass energy,
$t=s_{bb'} \simeq - |p_{b'\perp}|^2=- |q_{\perp}|^2$ is the momentum
transfer, with $q=p_b+p_{b'}$
the momentum of the gluon exchanged in the crossed channel, and
where $\lambda\,,\nu$ and $\rho$ denote the helicity of the lepton pair,
the quark pair and the gluons, respectively. Using the impact factor for
the gluon vertex, $C(b_g^+;{b'}_g^-) = p_{b'\perp}/ p_{b'\perp}^*$,
according to the conventions of Ref.~\cite{DelDuca:2000ha}\footnote{In
Ref.~\cite{DelDuca:2000ha} the generators of the group are normalised to
1/2, while in this paper to 1. We introduced the explicit $1/\sqrt{2}$
factor in \eqn{hecolpp} to take into account this difference.}
we obtain the impact factor for $e g^*\to e q \bar{q}$,
\bea
\lefteqn{V_f(a_\eb^-;{a'}_e^+,1_q^+,2_\qb^-)} \label{disifpp2}\\
&=& - {i\over k^2}
\left[{x (1-x)\over - x (1-x) k^2 + |p_{2\perp}|^2}
\left( p_{a\perp} \sqrt{p_2^+\over -p_a^+} + p_{2\perp} \sqrt{-p_a^+\over
p_2^+}\right) \left( p_{2\perp}^* \sqrt{p_{a'}^+\over p_1^+} +
p_{a'\perp}^* \sqrt{p_1^+\over p_{a'}^+} \right) \right. \nonumber\\
&& \left. \quad + \quad {x (1-x)\over - x (1-x) k^2 + |p_{1\perp}|^2}
\left( p_{1\perp}^* \sqrt{p_{a'}^+\over p_1^+} -
p_{a'\perp}^* \sqrt{p_1^+\over p_{a'}^+} \right)
\left( p_{a\perp} \sqrt{p_2^+\over -p_a^+} - p_{1\perp} \sqrt{-p_a^+\over
p_2^+}\right) \right]\, .\nonumber
\eea
Note that in \eqn{disifpp2} there is no divergence when the momenta $p_1$
and $p_2$ of the quark pair become collinear.
As a check of the calculation (or of high-energy factorisation), we have
evaluated also the amplitude (\ref{bdk6}) in the high-energy limit, and
have obtained the same result as in \eqn{hecolpp}, up to the substitution
\begin{equation}
i \gs\, f^{bb'c}\, C(b_g^{\nu_b};{b'}_g^{\nu_{b'}}) \leftrightarrow \gs\,
T^c_{b' \bar b}\, C(b_\qb^{-\nu_{b'}};{b'}_q^{\nu_{b'}})\,
,\label{qlrag}
\end{equation}
with impact factor for the quark vertex,
$C(b_\qb^+;{b'}_q^-) = (-\i)p_{b'\perp}/|p_{b'\perp}|$.

Using discrete symmetries of the helicity amplitudes, we find the
impact factors with the helicities of the quark and/or the lepton pairs
flipped,
\bea
&&
V_f(a_\eb^+;{a'}_e^-,1_q^-,2_\qb^+) = [V_f(a_\eb^-;{a'}_e^+,1_q^+,2_\qb^-)]^*\:,
\nn\\ &&
V_f(a_\eb^-;{a'}_e^+,1_q^-,2_\qb^+) = V_f(a_\eb^-;{a'}_e^+,2_q^+,1_\qb^-)\:,
\label{flipped}
\\ &&
V_f(a_\eb^+;{a'}_e^-;1_q^+,2_\qb^-) = [V_f(a_\eb^-;{a'}_e^+;2_q^+,1_\qb^-)]^*\:.
\nn
\eea
The impact factor for the backward kinematics is
\beq
V_b(a_\eb^-;{a'}_e^+;1_q^+,2_\qb^-) =
e^{\i \phi}\, V_f({a'}_\eb^-;a_e^+;2_q^+,1_\qb^-)|_{+ \to -} \:,
\label{backward}
\eeq
where the phase $e^{\i \phi}$ for us is immaterial since we compute
squared amplitudes and the index $+ \to -$ means that the plus components
in \eqn{disifpp2} are replaced with minus components.

\eqn{disifpp2} can be decomposed further in terms of an impact factor for
$\gamma^* g^* \to q \qb$ by factoring the lepton current
times the photon propagator
$\langle p_a |i\,\gamma^\mu| p_{a'} \rangle (-i g_{\mu\nu})/k^2$,
\beq
V_f(a_\eb^-;{a'}_e^+,1_q^+,2_\qb^-) =
\langle a- |\i \gamma^\mu| {a'}- \rangle \frac{-\i g_{\mu\nu}}{k^2}
V_{\gamma^*}^\nu(k;1_q^+,2_\qb^-)\:.\label{ifgamma}
\eeq
In light-cone notation, the lepton current is
\beq
\langle p_a- |i \,\gamma^\mu| p_{a'}- \rangle =
-2 \left(\sqrt{-p_a^+ p_{a'}^+},\,
\frac{-p_{a\perp} p_{{a'}\perp}^*}{\sqrt{-p_a^+ p_{a'}^+}};\,
-p_{a\perp} \sqrt{\frac{p_{a'}^+}{-p_a^+}},\,
p_{{a'}\perp}^* \sqrt{\frac{-p_a^+}{p_{a'}^+}}\right)\, ,\label{lepcurrent}
\eeq
with the $\gamma$ matrices chosen in the chiral representation
as in Ref.~\cite{DelDuca:2000ha}. The impact
factor $V_{\gamma^*}^\mu(k;1_q^+,2_\qb^-)$ for the
$\gamma^* g^* \to q^+ \qb^-$ process can be written as
\bea
\label{Vgamma-qqb}
\lefteqn{ V_{\gamma^*}^\mu(k;1_q^+,2_\qb^-) } \\ &=&
\frac{\sqrt{x (1-x)}}{-x (1-x) k^2+|p_{2\perp}|^2}
\left(x (1-x) k^+,\, \frac{|p_{2\perp}|^2}{-k^2} k^-;\,
x p_{2\perp},\, -(1-x) p^*_{2\perp}\right) \nn\\
&-& \frac{\sqrt{x (1-x)}}{-x (1-x) k^2+|p_{1\perp}|^2}
\left(x (1-x) k^+,\, \frac{|p_{1\perp}|^2}{-k^2} k^-;\,
-x p_{1\perp},\, (1-x) p^*_{1\perp}\right)\:.\nn
\eea
Using $k = (k^+,k^-;0_\perp)$, we can easily check that
$k_\mu V_{\gamma^*}^\mu(k;1_q^+,2_\qb^-) = 0$.

Finally, if we contract with the
polarization vector, $\vep^\mu = (\vep^+,\, \vep^-;\, \vep_\perp)$,
of the virtual photon, we obtain
\bea
\label{vepVgamma-qqb}
&&
V_{\gamma^*}^{\vep}(k;1_q^+,2_\qb^-)
\\ \nn && \quad
= \frac{\sqrt{x (1-x)}}{-x (1-x) k^2+|p_{2\perp}|^2}
\left(\sqrt{x (1-x)} k^+ \varepsilon^-
+\frac{|p_{2\perp}|^2}{-k^2} k^- \varepsilon^+
+(1-x) p_{2\perp}^*\varepsilon_\perp
-x p_{2\perp}\varepsilon_\perp^*\right)
\\ \nn && \quad
-\frac{\sqrt{x (1-x)}}{-x (1-x) k^2+|p_{1\perp}|^2}
\left(\sqrt{x (1-x)} k^+ \varepsilon^-
+\frac{|p_{1\perp}|^2}{-k^2} k^- \varepsilon^+
-(1-x) p_{1\perp}^*\varepsilon_\perp
+x p_{1\perp}\varepsilon_\perp^*\right)
\:.
\eea
We have checked that using this form of the impact factor, we can
reproduce the cross section in the high-energy limit obtained in
Ref.~\cite{Brodsky:1997sd}.

The impact factor with the helicity of the quark
pair flipped is obtained by exchanging the
momenta of the quark and antiquark,
\beq
V_{\gamma^*}^\mu(k;1_q^-,2_\qb^+) =
V_{\gamma^*}^\mu(k;2_q^+,1_\qb^-)\:,
\eeq
The impact factors for the backward kinematics, $k^- \gg k^+$, have the
same functional form as given by Eq.~(\ref{Vgamma-qqb}) (up to a phase)
with the $+$ and $-$ components as well as the quark helicities
interchanged.

As a final check, we computed the impact factor for $e g^*\to e q \bar{q}$,
\eqn{disifpp2}, also in the electron-parton frame, and verified that
it agrees with the colour-subleading piece, termed $B_2$, of the
impact factor for $q g^*\to q Q \bar{Q}$, computed in
Ref.~\cite{DelDuca:2000ha}.



\end{document}